\documentclass[10pt,preprint]{aastex}

\slugcomment{Accepted for publication in The Astrophysical Journal}

\shorttitle{Ages for field stars using gyrochronology}
\shortauthors{Barnes}

\begin{document}


\title{Ages for illustrative field stars using gyrochronology: viability, limitations and errors}


\author{Sydney A. Barnes}
\affil{Lowell Observatory, 1400 W. Mars Hill Rd., Flagstaff, AZ 86001, USA}
\email{barnes@lowell.edu}


\begin{abstract}
We here develop an improved way of using a rotating star as a clock, 
set it using the Sun, and demonstrate that it keeps time well.
This technique, called gyrochronology, permits the derivation of ages for 
solar- and late-type main sequence stars using only their rotation periods and 
colors.
The technique is clarified and developed here, and used to derive ages for
illustrative groups of nearby, late-type field stars with measured rotation 
periods. 
We first demonstrate the reality of the interface sequence, the unifying 
feature of the rotational observations of cluster and field stars that makes 
the technique possible, and extends it beyond the proposal of Skumanich by
specifying the mass dependence of rotation for these stars. 
We delineate which stars it cannot currently be used on.
We then calibrate the age dependence using the Sun.
The errors are propagated to understand their dependence on color and period.
Representative age errors associated with the technique are estimated at 
$\sim$15\% (plus possible systematic errors) for 
late\,F, G, K, \& early\,M stars.
Ages derived via gyrochronology for the Mt.\,Wilson stars are shown to be in 
good agreement with chromospheric ages for all but the bluest stars, and 
probably superior.
Gyro ages are then calculated for each of the active main sequence field stars
studied by Strassmeier and collaborators where other ages are not available.
These are shown to be mostly younger than 1\,Gyr, with a median age of 365\,Myr.
The sample of single, late-type main sequence field stars assembled by Pizzolato
and collaborators is then assessed, and shown to have gyro ages ranging from 
under 100\,Myr to several\,Gyr, and a median age of 1.2\,Gyr. 
Finally, we demonstrate that, in contrast to the other techniques, the 
individual components of the three wide binaries 
$\xi$BooAB, 61CygAB, \& $\alpha$CenAB yield substantially the same gyro ages.
\end{abstract}





\section{Introduction} \label{intro}

\subsection{Stellar age indicators, and motivation for a rotation clock}

The age of a star is its most fundamental attribute apart from its mass, and
usually provides the chronometer that permits the study of the time evolution
of astronomical phenomena. 
Consequently, a great deal of effort has been expended over the past several 
decades on the possibility of using stars as clocks, to reveal their own 
ages, those of the astronomical bodies associated with them, and to understand 
how various astronomical phenomena unfold over time.

The most successful of these chronometric techniques is the isochrone method 
(invented by Sandage, 1962; named and developed substantially by 
Demarque and Larson, 1964), 
based on the steady change in the color-magnitude morphology of a collection 
of stars, in response to the consumption and dimunition of their nuclear fuel
(e.g. VandenBerg et al., 2006; Demarque et al., 2004; 
Girardi et al., 2002; Kim et al., 2002; Yi et al., 2001). 

\subsubsection{The principal limitations of the Isochrone clock}

The isochrone technique fashions a collection of coeval stars of differing 
masses, i.e. 
a star cluster, into a remarkable clock that provides the age of the system.
However, vast numbers of stars, including our own Sun and most of the nearby 
stars amenable to detailed study, are no longer in identifiable 
clusters and spend their lives in relative isolation as field stars.
For these stars, the isochrone technique is less useful, because a star spends
most of its life burning hydrogen steadily on the main sequence, where its
luminosity and temperature, the primary indicators of isochrone age, are almost 
{\it constant}\footnote{For example, in the 4.5\,Gyr since it was on the zero 
age main sequence (ZAMS), the Sun's luminosity has increased by $\leq$50\% of 
its ZAMS value.}.
Using classical isochrones to tell the ages of single, low-mass, main sequence 
stars is akin to using gray hairs or baldness as an age 
indicator for toddlers, adolescents and adults!

Furthermore, the isochrone technique requires a measurement of the distance to
a field star to calculate its luminosity. This distance is hard to measure, and 
in fact, even after the publication of the results of the Hipparcos satellite 
(ESA, 1997), we know the distances to only $\sim$20,000 field stars 
(all of them nearby) to better than 10\% (Perryman et al. 1997). 
This imprecision leads to large errors in isochrone ages. 
A 10\% error on the distance to a solar twin would result in $\sim$20\% errors 
in its luminosity, and isochrone ages between 2 and 10\,Gyr\footnote{The Hipparcos satellite has indeed provided $\sim$1\% parallaxes for a group of stars, most of them bright enough to have been included in the catalog of Hipparchus himself if they were visible from Greece! To count them, you would need your own digits and those of some of your collaborators, but you wouldn't need more than a few of the latter.}.
Because the age of a star provides a direct link to many of its other 
properties, this deficiency is keenly felt. Knowledge of the age of a field 
star, however crude, is a very valuable astronomical commodity indeed. 

Thus we need to consider the possibility of fashioning clocks using other 
properties of (individual) stars. 
In particular, it would be very valuable to construct an age indicator that is 
independent of distance, and indeed, some of the activity-related indicators
suggested over the years, including the primary one used today, do have this 
valuable characteristic. 
In fact, the details of the pros and cons of the isochrone and other
chronometers are such that it might be useful here to step back even further 
and consider how an age indicator is constructed, 
and the general characteristics desirable for stellar age indicators.

\subsubsection{Steps in the construction of age indicators}

Five major steps seem to describe the process:
\begin{enumerate}
\item One needs to find an observable, $v$, that changes sensitively and
smoothly, perhaps monotonically, with age. Preferably, this observable would be 
a property of individual stars rather than that of a (co-eval) collection of 
them.
\item One needs to determine the ages of suitable calibrating objects 
independently. These would provide the connection to the fundamental units like
earth rotations, pendulum swings, etc.
\item One needs to identify and measure the functional form of the variable:
$v = v(t, w, x, ...)$ where $t$ is the age, and $w$, $x$, $...$, are possible
additional dependencies of the variable $v$. It is preferable to have fewer
variables, and to have separable dependencies of the form 
$v = T(t) \times W(w) \times X(x) ....$.
\item One needs to invert the dependence determined experimentally, numerically
or otherwise, to find $t = t (v, w, x, ...)$. This is usually non-linear, and 
sometimes has undesirable kinks.
\item Finally, one needs to calculate the error 
$\delta t = \delta t (t, v, w, x, ...)$.
\end{enumerate}

\subsubsection{Characteristics desired for age indicators}

The foregoing considerations suggest that the following characteristics are 
desirable for stellar age indicators.
\begin{enumerate}
\item {\bf Measurability for single stars: } 
      The indicator should be properly defined, measurable easily itself, and 
      preferably should not require many additional quantities to be measured,
      otherwise it cannot be used routinely.
\item {\bf Sensitivity to Age: } 
      The indicator should have a sensitive dependence on age, i.e., should
      change substantially (and preferably regularly) with age,
      otherwise the errors will be inherently large.
\item {\bf Insensitivity to other parameters: } 
      The indicator should have insensitive (or separable) dependencies on other
      parameters that affect the measured quantity,
      otherwise there is the potential for ambiguity.
\item {\bf Calibration: }
      The technique should be calibrable using an object (or set of objects)
      whose age(s) we know very well, otherwise systematic errors will be
      introduced.
\item {\bf Invertibility: } The functional dependence determined above should
      be properly invertible to yield the age as a function of the measured
      variables.
\item {\bf Error analysis: } 
      The errors on the age derived using the technique ought to be calculable,
      otherwise no confidence can be attached to the ages.
\item {\bf Test of coeval stars: }
      The technique should yield the same ages for stars expected to be coeval,
      otherwise the validity of the technique itself must be questioned.
\end{enumerate}
We summarize in Table\,1 how (in)adequately these characteristics are satisfied 
by the three age indicators relevant to this paper. 
While the entries, especially for gyrochronology, anticipate the results derived
in this paper, the characteristics desired guide the progress of, and form a 
continuous backdrop to this work.


\clearpage

\begin{deluxetable}{llll}
\tabletypesize{\scriptsize}
\tablecaption{Characteristics of the three major age indicators for field stars
\label{tbl-1}}
\tablewidth{0pt}
\tablehead{
\colhead{Property} & \colhead{Isochrone age}   & \colhead{Chromospheric age} & \colhead{Gyrochronology}   
}
\startdata
Measurable easily? & ?\tablenotemark{a} (Distance reqd.)& ?\tablenotemark{b} (Repetition reqd.) & ?\tablenotemark{c} (Repetition reqd.)\\
Sensitive to age? &  No (on MS)	& Yes                  & Yes \\
Insensitive to other parameters? &  No	   & Yes\tablenotemark{d} & Yes \\
Technique calibrable? & Yes (Sun) & ?\tablenotemark{e} (Sun?)  & Yes (Sun) \\
Invertible easily? & No & Yes & Yes \\ 
Errors calculable/provided?   &  ?\tablenotemark{f} (Difficult) & Yes?\tablenotemark{g} & Yes \\
Coeval stars yield the same age? & No (Field binaries) & ?\tablenotemark{h} & Yes \\
\enddata


\tablenotetext{a}{A field star requires a good distance measurement in order to
determine its luminosity for comparison with isochrones. As explained in the 
text, good distances are available to only a few such stars.}
\tablenotetext{b}{Another reason for this `?' is that it is not clear to an 
innocent bystander how to transform 
between the various quantities published as a chromospheric flux: 
$S$, HK\,index, $R$, or $R'_{HK}$. The lack of a defined standard quantity for
published work is a significant drawback.}
\tablenotetext{c}{Another reason for this `?' is that for old stars the 
modulation in broadband photometric filters is too small to yield a rotation
period, and for these stars one must resort to more onerous means such as
detecting the rotational modulation in the Ca\,II\,H\,\&\,K emission cores.}
\tablenotetext{d}{The benefit of doubt has been given but in fact, there is 
usually some black magic in the transformation between chromospheric flux and
age.}
\tablenotetext{e}{The relationship between chromospheric emission and age in 
Soderblom et al. (1991) is calibrated against isochrone ages of three 
``fundamental'' points, and those of the evolved components of visual binaries.
Since all isochrones are calibrated using the age of the Sun, this calibration
is also ultimately based on the age of the Sun, except for the additional step
involved.}
\tablenotetext{f}{Errors on isochrone ages for field stars were essentially 
non-existent until Pont \& Eyer (2004) suggested a Bayesian scheme that allows
one to determine whether or not an isochrone age is `well-defined' (Jorgensen 
\& Lindegren 2005) i.e. whether or not the probability density distribution for 
the age has an identifiable maximum, and if so, to calculate an error based on 
this property. This method has since been used by Takeda et al. (2007) on their 
field star sample.}
\tablenotetext{g}{Soderblom et al. (1991) provide the error on their fit, in 
this case $\sim$0.17\,dex ($\sim$40\%), of chromospheric emission to 
(isochrone) age for their sample. Other researchers, 
including Donahue (1998), usually do not provide errors.}
\tablenotetext{h}{For the eight pairs in Table\,2 of Donahue (1998), the mean
discrepancy is 0.85\,Gyr for a sample with a mean age of 1.85\,Gyr, so that the
fractional discrepancy in age is 0.46, or just under 50\%.}


\end{deluxetable}

\subsubsection{Motivations for investigating a rotational clock}

A wide array of age indicators have been developed over the past decades.
The most well-known are chromospheric emission (Wilson, 1963) and 
rotation (Skumanich, 1972), but others like surface lithium abundance 
(Vauclair, 1972; Rebolo, Martin \& Maguzzu 1992; Basri, Marcy \& Graham 1996; 
Stauffer 2000) and  coronal emission in X-rays (Kunte et al. 1988), usually 
through its dependence on rotation (Pallavicini et al. 1981; Gudel 2004),
have also occasionally been suggested and used in various contexts.
All of these are related to stellar activity and are based on empirical 
correlations between the property in question and stellar age. 
They have been considered less reliable clocks than the canonical isochrone 
technique because the underlying physics is not well understood, and in fact 
there is a great deal of debate even about what the important underlying 
phenomena are. 
Finally, one must also consider whether and how each of these age indicators is 
calibrated.

Ever since the work of Skumanich (1972), and especially since the work of 
Noyes et al. (1984), the relationship between chromospheric emission and age 
has enjoyed the distinction of being the most consistent,
making chromospheric emission the leading age indicator for nearby field stars 
(e.g. Soderblom 1985; Henry et al. 1996; Wright et al. 2004).

But there are more fundamental stellar observables than chromospheric
emission. In fact, of all the activity-related properties of stars, rotation 
is undoubtedly the most fundamental, and many believe that together with 
stellar mass (and another variable or two), it might be responsible, directly or
indirectly, for the observed morphology of all the other activity indicators. 

In fact, besides being obviously independent of the distance to the star, 
stellar rotation is now known to change systematically, even predictably, 
on the main sequence, where the isochrone technique is at its weakest. 
Furthermore, the specific form of the rotational spindown of stars is such that 
{\it initial variations in the rotation rates of young stars appear to become 
increasingly unimportant with the passage of time, leading to an almost unique 
relationship between rotation period and age for a star of a given mass}.
Finally, rotation is a property we can now measure to great precision; 
rotation periods for late-type stars are sometimes determined today to better 
than one part in ten thousand\footnote{The usefulness of this precision is less clear in the context of the differential rotation with latitude of the Sun and solar-type stars, but it is also clear that we are beginning to understand the systematics and origin of differential rotation, so that the attainment of such precision is useful in other ways as well.}! 
These features of stellar rotation - its predictability, measurability, and
simplicity - suggest that some effort is warranted in improving its use
as an age indicator beyond the relationship suggested by Skumanich (1972).

In fact, as we shall show below, and as is summarized in Table\,1, 
gyrochronology satisfies more of the criteria required for an age indicator as 
listed above, than any other 
astronomical clock in use, and appears to be complementary to the isochrone 
technique, in that it works very well on the main sequence, while the isochrone 
method is better suited to evolved stars.

This paper addresses the issues of constructing and calibrating a rotational
clock. It appears that to first order stellar rotation depends only on the mass 
and age of the star, so that jointly taking account of these dependencies of 
rotation permits the determination of rotational ages ({\it and their errors}) 
for a substantial sample of main sequence stars, and 
even {\it individual field stars}, a technique we suggest be called 
``gyrochronology.''

\subsection{Stellar rotation as an astronomical clock}

Major steps in the direction of using stellar rotation as a clock were made by 
a series of studies in the 1960s, culminating in the famous relationship of 
Skumanich (1972), relating the averaged surface rotational velocities, 
$\overline{v \sin i}$, of stars in a number of open clusters to their ages, $t$, 
via the expression: $\overline{v \sin i} \propto 1/\sqrt{t}$. 
Skumanich noted that the equatorial surface rotation velocity of the Sun 
at its independently derived age also matched this relationship\footnote{The 
age of the Sun is not directly known, of course. We use the age of the formation
of the refractory inclusions in the Allende meteorite as an estimate of the
Sun's age (e.g. Allegre et al. 1995 but see also Patterson 1953; 1955; 1956; 
Patterson et al. 1995 and Murthy \& Patterson, 1962 for the original work 
establishing that the age of the Earth and that of the meteorites is identical 
and can be called the ``age of the solar system'').}. 
Over the years, astronomers have come to believe that this relationship 
encapsulates something fundamental about the nature of winds and angular 
momentum loss from late-type stars\footnote{It appears to be equivalent to a 
cubic dependence on the rotation speed, $\Omega$, of the angular momentum loss 
rate, $dJ/dt$, from solar- and late-type stars: $ dJ/dt \propto -\Omega^3$ 
(Kawaler, 1988). In fact, parameterizations based on this behavior are 
routinely incorporated into stellar models that include rotation (e.g. 
Pinsonneault et al 1989). Two of these three powers of $\Omega$ appear to be 
related to the strength of the magnetic field of the star under the assumption 
of a linear dynamo.}.

Skumanich (1972), however, did not specify the mass-dependence of rotation 
- the so-called `correction for color' that he performed. Presumably this 
correction was based on the Kraft (1967) curve or something similar.
There is also a measurement issue - for individual stars the ambiguity inherent 
in using $v \sin i$ measurements, with the generally unknown angles of 
inclination, $i$, can be expected to introduce large errors in the age 
determinations\footnote{Projection effects are less relevant for entire 
clusters, as with the averaged $v \sin i$ measurements that Skumanich used. 
Presumably they average out because they are similar from cluster to cluster.}. 

Furthermore, mass-dependent comparisons of rotation require precise values for 
stellar radii to infer the true angular rotation speeds of stars.
Despite these shortcomings, various studies have occasionally used this 
relationship for rotational ages, e.g. Lachaume et al (1999), and the ages 
derived in this manner are in rough agreement with ages derived 
using other techniques, but they are not noticeably better. 

Kawaler (1989) attempted an empirical color correction using a linear function
of the $(B-V)$ color of the star, but he provided no physical basis for such a 
correction - indeed, there is none - and in any case it breaks down dramatically
for late-F to early-G stars (see especially Fig.\,1 in his paper). The specific 
ways in which stars of different masses spin down, whether young clusters obey 
such a spindown or not, and how observations in young clusters are related to 
field star observations is a continuing matter of debate and discussion.

If it were possible to eliminate the ambiguity in $v \sin i$ observations
by finding the true angular rotation rates of stars, as is routinely
accomplished nowadays by measuring rotation periods\footnote{Rotation periods are measured by timing the modulation of either filtered starlight, which works well for young stars (e.g. Van Leeuwen et al. 1987), or that of the chromospheric emission (e.g. Noyes et al. 1984), which works for older stars. Either of these is obviously more demanding than deriving $v \sin i$, but the effort is well worth the results, and furthermore, is being done routinely, as detailed below. As an aside, we point out that the ``rotation periods'' listed by Wright et al. (2004) are {\it not} directly measured; they are calculated from the measured chromospheric emission, and hence unsuitable for our purposes.}, 
and if the periods were to have a unique and ``correctable'' dependence on 
color, with reasonably small scatter, rotation could become incredibly useful
as a stellar clock.

Using the (measured) rotation periods of the Mt.\,Wilson stars, Barnes (2001) 
showed that 
the age dependence of rotation for these stars is indeed Skumanich-type 
($ P \propto \sqrt{t}$), and furthermore, the mass dependence of rotation for 
these stars is similar to that observed in the Hyades open cluster. 
Barnes (2003a) noted that an age-increasing fraction of open cluster
stars and essentially all solar-type stars beyond a few hundred Myr in age,
including individual field stars, obeyed the same mass dependence. These two 
facts provide the connection between rotation in clusters and in the field.

Furthermore, Barnes (2003a) wrote down this mass dependence, $f$, as a 
convenient function\footnote{He used the the function: $f(B-V) = \sqrt{B-V-0.5} - 0.15 (B-V-0.5)$ but $f$ can of course be written in terms of any convenient function of stellar mass. We will modify the expression for $f$ below.} of $(B-V)$ color, $f(B-V)$. 
This function, $f$, appears to be closely related to the moment of inertia, 
$I_*$, of the entire star via $f \propto 1/\sqrt{I_*}$. 
This identification, the rotational implications for the Sun and cluster stars, 
and for stellar magnetic fields, are discussed at length in Barnes (2003a and 
2003b), but here we are concerned only with the 
{\it universality and uniqueness} of this function, apparently separable from 
the age dependence, a circumstance that leads to a remarkably simple way of 
deriving ages (and their errors) for solar-type stars on the main sequence\footnote{I have learned from Ed Guinan (2006, personal communication) that he has been using the Hyades rotational sequence and the Skumanich relation to derive stellar ages. That would make it substantially similar to the technique developed here.}.

\subsection{Proximate motivations for constructing a rotational clock}

There are also proximate motivations for this work. It has become 
increasingly obvious that greater precision in stellar ages than is available
using isochrones and chromospheric emission is required for many astronomical
purposes. The effort currently being expended on the host stars of planetary
systems is a case in point. Well determined ages would eventually permit the
study of the the evolution of planetary systems. This application is a proximate
one relevant to our time, but the method can undoubtedly be used to tackle
some of the deeper problems in astronomy.

The requirement of a stellar rotation period is not as onerous as might
initially appear\footnote{We note here that chromospheric emission measurements
also require repeated measurement to ensure that they are averages over the 
variability from rotation or from stellar cycles.}. 
{\it As opposed to the requirement for isochrones, it avoids the necessity of deriving the distance to a field star.}
The Vanderbilt/Tennessee State robotic photometric telescopes (e.g. Henry et 
al. 1995) and of the University of Vienna (e.g. Strassmeier et al. 2000) in 
Southern Arizona are designed to derive stellar rotation periods, and in fact, 
the Strassmeier group, now in Potsdam, has almost finished the construction of
two 1.2m telescopes, Stella\,1\,\&\,2, to monitor active stars almost 
exclusively (Strassmeier, 2006). 
The ASAS project (e.g. Pojmanski, 2001) routinely monitors and catalogs stellar 
(and other) variability in the Southern hemisphere, and a Northern counterpart  
is the Northern Sky Variability Survey (Wozniak et al. 2004). 

The Canadian MOST satellite (Matthews et al. 2000) was launched to provide 
(and has since delivered) superb time-series photometry  (witness its 
identification of two closely spaced rotation periods for $\kappa^1$\,Ceti, 
corresponding to two spot groups; (Rucinski et al. 2004), its detection of 
0.03\%-0.06\% brightness variations in a subdwarf B star; (Randall et al. 2005),
and its recent identification of g-modes in $\beta$ C\,Mi (Saio et al. 2007)). 
The COROT satellite mission has been designed\footnote{In fact, the satellite has been built and launched.} to study stellar convection, rotation, and now, planetary transits. 
A number of ground-based telescopes are planning to or already exploiting the
time domain, and of these the Large Synoptic Survey Telescope (LSST) 
perhaps has the greatest visibility.

The Kepler space mission, being readied for launch, is likely to yield not 
only the planetary transit, but also the rotation period of the host star. 
In fact, Kepler is likely to yield rotation periods for orders of magnitude 
more stars than planetary transits\footnote{Assuming that we do not throw the baby out with the bath water.}. Regardless of whether or not the Kepler mission 
delivers what it promises, stellar rotation periods will be determined 
routinely as time domain astronomy comes into its own. 
A very significant portion of time-domain work on stars will yield the stellar
rotation period (it is a by-product of all searches for planetary transits), 
and if this measurement can be used to derive a precise stellar age, it would
permit us to address many problems involving chronometry that are not presently
solvable.

\subsection{Overview of the paper and sequence of succeeding sections}
Our goal here is to specify the stars for which gyrochronology can and cannot 
be used, to develop it to yield useful ages for individual field solar-type 
stars, and to calculate the errors on these ages. 
We will also show that where both are available, these new ages agree with
(and might even supercede) the ages provided by other methods.

We begin by showing that rotating stars, whether in clusters or
in the field, are of two types\footnote{In principle, there is a third type, g,
representing stars in transition from the first/C to the second/I type.}: 
fast/Convective/C and slow/Interface/I. 
The Sun is shown to be on the interface sequence, which defines the rotational
connection between all solar-type stars (section\,2). 
These stars are shown to spin down Skumanich-style, 
with a mass dependence that 
is shown to be universal, and for which we derive a simple functional form using
stars in open clusters (section\,3). 
These functional dependencies are combined to yield a simple expression for the
gyro ages of stars.

In section\,4, we derive the errors on these ages.
Section\,5 demonstrates that these ages compare favorably with chromospheric 
ages for a well-studied sample.
Sections\,6 and 7 illustrate the use of gyrochronology on samples for which 
other ages are not uniformly available, namely the field star samples of 
Strassmeier et al. (2000) and Pizzolato et al. (2003). 
Section\,8 demonstrates that gyrochronology yields the same age for the two
component stars of wide binaries.
Section\,9 contains a comparison to recently derived isochrone ages for a common
subset of the stars considered here, 
and Section\,10 contains the conclusions.

\section{The rotational connection between all solar-type stars}


A fundamental fact of stellar rotation is that there are two major varieties,
C \& I, of rotating solar-type (FGKM) stars (see Barnes 2003a). 
A third variety, g,  merely represents stars making an apparently 
unidirectional transition from one variety (C) to the other (I).
All three varieties of stars are normally found in young open clusters, 
but the Sun and all old solar-type stars are of only the I variety.
Each of these varieties of rotating stars has separate mass- and age 
dependencies
that can be clarified considerably merely by effecting the correct
separation of the stars by variety. One of these, called the Interface (I) 
sequence stars, containing the Sun and all old field solar-type stars, 
is related to the property Skumanich noticed in 1972.
This group is the one that we will use here to demonstrate the technique
of gyrochronology, because stars change into this variety over time.
We are fortunate that the rotational mass- and age dependencies of this
group of stars appear to be both separable and also particularly simple.

If the mass- and age dependencies of this sequence are indeed separable,
as was claimed by Barnes (2003a) to be of the form 
	$ P (t, M) = g(t) . f(M) $, 
then merely dividing the measured rotation periods $ P (t, M) $ by the
functional form $ g(t) $ of the age dependence should make the mass
dependence $f(M)$ manifestly clear. 
For observational convenience, and also to avoid the error inherent in the
conversion from $B-V$ to stellar mass, we have used $f(B-V)$ instead of $f(M)$. 
Removing an assumed Skumanich-type age dependence, where $ g(t) = \sqrt{t} $,
is particularly simple, and appears to bring the I sequence into sharp focus,
leading to the identification of the mass dependence as a function 
$ f = f(B-V) $.
In the two subsections below, we effect this determination separately for
cluster and field stars.

\subsection{The connection between clusters themselves}

Here we show that $f$ represents the connection between most rotating stars, 
and that the functional dependence 
of $f$ on color or stellar mass is common to all open clusters.
We use all the open cluster rotation periods currently available in the
literature; note that we are restricted to those stars for which $(B-V)$ 
colors are also available. The major sources are listed in Table 2.
We divide each of the measured rotation periods by $g(t)=\sqrt{t}$
where $t$ is the age of the cluster in Myr, as listed in Table\,2.
These quantities are plotted against de-reddened $(B-V)$ color
in Fig.\,1 , on a linear scale in the upper panel, and on a logarithmic
scale in the lower panel.
Clusters are color-coded violet through red in increasing age sequence.


\clearpage

\begin{deluxetable}{lrl}
\tabletypesize{\scriptsize}
\tablecaption{Principal sources for open cluster rotation periods. \label{tbl-2}}
\tablewidth{0pt}
\tablehead{
\colhead{Cluster} & \colhead{Age}   & \colhead{Rotation Period Source}   
}
\startdata
	IC 2391	 &  30\,Myr	& Patten \& Simon (1996)\\
	IC 2602	 &  30\,Myr	& Barnes et al. (1999)\\
	IC 4665	 &  50\,Myr	& Allain et al. (1996)\\
       Alpha\,Per &  50\,Myr	& Prosser \& Grankin (1997)\\
	Pleiades & 100\,Myr	& Van Leeuwen, Alphenaar \& Meys (1987), 
					Krishnamurthi et al. (1998)\\
        NGC\,2516& 150\,Myr     & Barnes \& Sofia (1998)\\
	M\,34	 & 200\,Myr	& Barnes (2003a)\\
	NGC\,3532& 300\,Myr	& Barnes (1998)\\
	Hyades	 & 600\,Myr	& Radick et al. (1987)\\
	Coma	 & 600\,Myr	& Radick, Skiff \& Lockwood (1990)\\
\enddata




\end{deluxetable}

The most striking aspect of these data is the curvilinear feature representing
a concentration of stars in the vicinity of the solid line. 
This is the interface sequence, I, proposed in Barnes (2003a), the one that
will consume our attention in this paper,
and whose position we will use as an age indicator for field stars.
Along the bottom of the upper panel one may also discern another linear
concentration of stars which represents the convective sequences, C, of the
youngest open clusters. This sequence could also potentially be used as
an age indicator for young stars, but its dependencies on stellar age
and mass are more complicated than those of the I sequence (see Barnes 2003a),
and we do not use it here.
Stars located between these sequences are either on the convective sequences 
of the older open clusters in this sample, or in the rotational gap, g, between 
the interface and convective sequences.

Every single cluster plotted in Fig.\,1 possesses an identifiable  interface 
sequence.
The fraction of stars on this sequence increases systematically with cluster 
age, as shown earlier in Barnes (2003a); see especially Fig.\,3 there. 
But for us the crucial feature of these data is that these age-corrected 
sequences overlie one another. This feature is shared by all open
clusters, and can be represented by a function $f(B-V)$, common to
all clusters. A particular choice (used in Barnes 2003a) of 
$f(B-V): \sqrt{(B-V-0.5)} - 0.15 (B-V-0.5)$ is displayed in both panels. 
This is of the nature of a trial function, useful in locating the I sequence
roughly, and we will improve on this choice subsequently. 

Barnes (2003a) has suggested that $f$ ought to be identified with 
$1/\sqrt{I_*}$, where $I_*$ is the moment of inertia of the star, implying a 
substantial mechanical coupling of the entire star on this sequence. 
The suggestion in that publication was magnetic coupling by an interface 
dynamo, hence the name interface sequence for this group of stars.

The dotted lines in the figure are drawn at $2 f$, and $4 f$. 
Present indications are that some of these stars are either non-members of 
the cluster, sometimes stars with spurious/alias periods or otherwise
misidentified variables of another sort.
Note that the Hyades, where excellent membership information is available, 
has no stars above the sequence.
A similar situation obtains in NGC\,2516 and M\,34, which are also relatively 
clean samples. Good cluster membership information could resolve this issue 
completely.

In summary, the behavior of the open cluster rotation observations suggests
the existence of a feature common to all open clusters, the Interface sequence,
which is observationally definable by its common mass
dependence, $f(M)$, across clusters, here represented by $f(B-V)$.
These observations also justify the use of the Skumanich (1972) relationship 
between rotation and age to describe the age dependence of rotation, but 
{\it only for rotating stars of this particular (interface) type}.

\subsection{The connection between clusters and field stars}

Here we show that the mass dependence, $f$, among open clusters, is also
shared by field stars as exemplified by the Mt.\,Wilson stars.
We begin by removing from the Mt.\,Wilson sample those stars known or suspected
not to be dwarfs (based on Baliunas et al. 1995 
), to avoid any possible complexity 
related to structural evolution off the main sequence. 
An effective connection with open clusters requires splitting the remaining 
main sequence Mt.\,Wilson field star sample by age, to control the age 
variation among the stars and gain leverage over the time domain. 
Fortunately, one such split, based on detailed studies of this sample, 
especially of chromospheric emission, has already been made by Vaughan (1980),
who classified these stars into a Young (Y) and Old (O) group.
The simplest course of action is to use the existing divisions.
Although the age divisions are quite broad, subsequent work has confirmed
the basic classification. A cut by chromospheric activity is well-known
to be also a cut by rotation and age 
(e.g. Barnes 2001, and references therein).
Part of the goal of this paper is to develop a way of ordering the stars by 
age, so we cannot start by assuming chromospheric ages for individual stars. 

As a result of the above classification, we have two groups of stars, Y and O,
consisting of 43 and 49 stars respectively, equivalent to two additional open 
clusters, each containing stars with a wide range of ages. 
What are these ages? Barnes (2001) [see especially Fig.\,3] and Barnes (2003a) 
[see especially Fig.\,2] suggest that, in terms of rotation,
the young (Y) stars range in age from less than 300\,Myr to about 2\,Gyr,
with a characteristic age of 800\,Myr, while the old (O) stars range in age
from 2\,Gyr to about 10\,Gyr, with a characteristic age of 4.5\,Gyr.
The age of the Y group is older than, but comparable to, young open cluster 
ages, 
while the age of the O group is reasonably represented by the Sun's age. 
Effectively, we are assuming that the Sun is an appropriate representative,
in rotation and age, of the Old Mt.\,Wilson sample. 

If we use the chromospheric ages for the same Y and O groups of Mt.\,Wilson
stars, calculated using the relationship of Donahue (1998), 
the median ages of the same samples work out to be 780\,Myr and 4.24\,Gyr 
respectively, reasonably close to our assumption above. 
We will use these new values as the representative ages for the Y and O groups
in this paper. 
The rotation clock can easily be recalibrated when the need arises.

We can make the field star data comparable with the open cluster data by
similarly removing this approximate age dependence. Thus, we divide the
rotation periods of the Young Mt.\,Wilson stars by $\sqrt{t_Y=780\,Myr}$
and display them using small black asterisks in Fig.\,2, overplotted on the
open cluster data (colored circles). Similarly, we divide the rotation
periods of the Old Mt.\,Wilson stars by $\sqrt{t_O=4240\,Myr}$ and display
them in Fig.\,3 using large black asterisks, again overplotted on the open 
cluster data.

Examination of Figs. 2 and 3 leads to several conclusions.
Firstly, we note that both the Young and Old samples overlie the Interface
sequences of the open clusters. The greater dispersion of the Y and O stars
relative to those of the open cluster I sequences can be traced to the age
dispersion in each of these samples. We can see, despite this dispersion,
that C sequence stars, and possibly g (gap) stars, are absent from both the
Young and Old samples. These data are consistent with all of the Mt.\,Wilson
stars being of the I variety.

Any doubts about the classification of the Young and Old Mt.\,Wilson stars can 
be settled by making individual corrections for these stars based on their
chromospheric ages. If these ages are correct, then removing their dependence,
as in the open clusters, should make the mass dependence obvious, and that
mass dependence ought to be similar to $f$.
In fig.\,4 we display the result of dividing the Mt.\,Wilson star rotation
periods by the square root of the (individual) chromospheric ages, calculated 
using the formula from Donahue (1998)\footnote{This formula yields ages in close agreement with those for old stars calculated using the formulae in Soderblom, Duncan \& Johnson (1991),  but is generally considered to be an improvement for young stars because saturation effects are taken into account [cf. Barnes 2001].}.
We note that almost all of the Mt.\,Wilson stars in the color range considered 
lie on/near $f(B-V)$.
Fig.\,4 displays $f(B-V)$ (solid line), $0.8f$ and $1.25f$ (dotted lines), to
show this proximity.
Indeed, a free-hand fit would be almost identical to $f$.
We will improve on this trial function below.
It is likely that this sample does not contain any C sequence stars or even any
gap stars.
(This observation is consistent with the C$\rightarrow$I transition timescale
of $\sim$200\,Myr observed in open clusters.)

In closing this section, we reiterate that the rotation period distributions
of open clusters, when corrected for a Skumanich-type age dependence, display
two strong concentrations of stars. The slower of these consists of sequences
that are common to all open clusters and overlie one another.
Rotation period distributions of main sequence field stars, despite the
difficulty of correcting for their ages, also display this same sequence.
The other concentration of stars, present in open clusters, is absent here.
The feature common to cluster and field stars, called the Interface sequence,
can be fit by a function (as we do below), 
giving the mass dependence of stellar rotation.



\section{(Re-)determination of the mass- and age dependencies}

Having determined that rotation has both mass- and age dependencies, how is one
to specify them independently using one set of data, and without greatly 
compromising the determined dependencies?
One way forward is to realize that open clusters can specify the mass dependence
regardless of whether or not we make some error in their ages - after all, they
are all clustered near ZAMS ages - while the Sun provides a datum with a very
well-defined age far out, but obviously no information about the mass 
dependence. These facts suggest the use of open clusters to decide the 
mass dependence, and the Sun to decide the age dependence. 
The effect is to follow the Copernican Principle and assume that the Sun is 
the perfect representative of its class of star. 
(We note that the same principle guides the solar calibration of the 
classical isochrone method.)

The construction of an appropriate fit for the mass dependence requires the
removal of stars that are not on the I\,sequence in open clusters.
This cannot yet be done unambiguously using only color-period data because the
position of the I\,sequence has yet to be specified well. 
That is part of the goal of this paper.
For clusters where X-ray data are also available, we get a additional handle on
classifying these stars using the correspondence noted in Barnes (2003b). 
There, the classification in X-rays of unsaturated, saturated, and 
super-saturated stars is shown to correspond, on a star-by-star basis with 
I-, g-, and C\,stars respectively. 
We therefore select the unsaturated stars, which are all I\,sequence stars in 
the color-period diagram, and for each cluster, we plot $P/\sqrt{Cluster\,Age}$ 
against $(B-V)_0$. These stars define a sequence in  color-$P/\sqrt{Age}$ space,
and we can now discard stars from the other clusters without X-ray information 
that lie far away from this sequence. The aim is to do this conservatively, so
as to retain as many stars as possible for a proper definition of the 
I\,sequence, while removing clear C-, g-, or alias period stars.
While it is true that this determination is done subjectively at present, it is
done as empirically as we possibly can at the present time\footnote{Judgements such as these are routinely made during classical isochrone fitting. A rich dataset or two, such as the one for M35 (S. Meibom, in prep) should eliminate much of the ambiguity within a year or two.}. 
The remaining stars happen to lie near the trial function $f(B-V)$ that was 
used in Barnes (2003a), but this function does not obviously pre-determine the 
new one.

This exercise suggests that slight modifications to the open cluster ages are 
needed to tighten the overlap of the individual I sequences. We have made these 
slight adjustments in order to ensure a valid result for the mass dependence.
The ages used are:
IC\,4665:    40Myr;
Alpha\,Per: 110Myr;
Pleiades:   120Myr;
NGC\,2516:  180Myr;
M\,34:      200Myr;
NGC\,3532:  250Myr;
Coma\,Ber:  600Myr;
Hyades:     600Myr.
We have not used IC\,2391 and IC\,2602 because although they possess 
identifiable sequences, they are some distance off the sequence defined by
the other clusters, a fact we attribute to the residual effects of pre-main
sequence evolution\footnote{The 110\,Myr age for Alpha\,Per might also be a surprise to some. In fact, we guess that the underlying rotational behavior might also originate in residual effects from pre-main sequence evolution, similar to IC\,2391 \& IC\,2602. However, we have chosen to retain it in this analysis because we cannot yet afford to lose the many periods in this cluster (contributed by Prosser \& Grankin 1997).}. 
These minor age adjustments are justifiable because, in any case, we are not 
using the open clusters to decide the age dependence of rotation. We are 
effectively merely using them to set the ``zero-point'' of the age dependence.
We know that their I sequence age dependence is roughly Skumanich-style.

Having removed the non-I sequence stars, we note a tight mass-dependence for
which we desire a functional form. The trial function 
$f(B-V)= \sqrt{(B-V-0.5)} - 0.15(B-V-0.5)$ has an undesirable singularity at 
$(B-V)=0.5$, which we would like to move blueward, to accommodate the late F 
stars. We would also like to retain an analytic function.
A function of the form $f(B-V)=a.(B-V - 0.4)^{b}$ where $a$ and $b$ are
fitted constants, seems to be appropriate (and will permit appropriate error 
analysis later).
Using the R statistics package (Ihaka \& Gentleman, 1996) to do the fit, we get
$a=0.7725 \pm 0.011 $ and $b=0.601 \pm 0.024 $.
This function is plotted with a solid line in Fig.\,5 over the I sequence 
stars in the open clusters listed above. The standard error on the residuals
is $0.0795$ on 182 degrees of freedom.

To show that the fit is appropriate, we also display using a dashed line in
Fig.\,5 the result of fitting a non-parametric trend curve using the function
{\it lowess} in the R statistics package\footnote{The {\it lowess} function implements a locally weighted regression smoothing procedure using a polynomial. No significant difference is seen with other smoothing procedures.}. 
The close correspondence between the 
two curves shows that the function chosen above is appropriate for these data.

Having determined the mass-dependence using open clusters, we check that it is
appropriate for the field stars, which provided the motivation for improving
the representation of $f(B-V)$. We plot the new and old dependencies in 
Fig.\,6, over the the Mt.\,Wilson stars (same data as in Fig.\,4), and again 
{\it assuming that the chromospheric ages are correct}. 
The figure displays the difference between the old and new functions, $f$, in
relation to the Mt.\,Wilson stars. (This discrepancy between $f$ and the 
F\,star data is partially attributable to the assumption of correct 
chromospheric ages for blue stars and is addressed in another section below.)

Having specified the mass dependence using open clusters, and having shown that
the Sun and field stars also follow this mass dependence, we can now determine 
the age dependence. We know that the age dependence $g(t)$ will roughly be 
$\sqrt t$, but the open clusters are too young to be effective calibrators, 
nor are their ages known to sufficient precision. 
In contrast, the rotation rate of the Sun is perhaps the most fundamental datum 
in stellar rotation, and its parameters are the fundamental calibrators for 
theoretical stellar models. 
In keeping with this tradition, (and older ones of calibrating clocks by the
sun), we choose to specify the age dependence via a solar calibration.
Representing the age dependence using $g(t) = t^n$, and calibrating the index 
$n$ using the Sun's measured mean rotation period of 26.09d 
(Donahue et al. 1996), a solar $B-V$ color of 0.642 (Holmberg et al. 2005) and 
a solar age of 4.566 Gyr (Allegre et al. 1995) yields $n=0.5189 \pm 0.0070$, 
where the error on $n$ has been calculated by simply propagating the errors on 
the other terms and assuming 1\,d and 50\,Myr errors in the period and age of 
the sun respectively. This calculation is detailed in the appendix to this
paper.

So, the final result works out to be:
$P(B-V,t) = f(B-V) . g(t)$, where
\begin{equation}
{
f(B-V) = (0.7725 \pm 0.011) \times (B-V_0 - 0.4)^{0.601 \pm 0.024}
}
\end{equation} 
and 
\begin{equation}
{
g(t)= t^{0.5189 \pm 0.0070}
}
\end{equation}
a result which is simultaneously analytical, simple, separable, almost 
Skumanich, fits the mass dependence of the open clusters, and the age 
dependence specified by the Sun. 

We know that for the I sequence stars, whether in clusters or in the field,
the rotation rate is given by $P(B-V,t) = f(B-V).g(t)$, where $f(B-V)$ and
$g(t)$ were determined above. This is true for each star.
Therefore $t = g^{-1}[P(B-V,t)/f(B-V)]$. 
Explicitly, 
\begin{equation}
log(t_{gyro}) = \frac{1}{n}\{log(P)-log(a)-b \times log(B-V-0.4)\}
\end{equation}
where $t$ is in Myr,  $B-V$ and $P$ are the measured color and rotation period 
(in days) respectively, 
$n=0.5189 \pm 0.007$, $a=0.7725 \pm 0.011$, and $b=0.601 \pm 0.024$.


\section{Errors in the ages}

The ages from gyrochronology become truly useful only when we can estimate their
errors and show that they are acceptable. 
A crude estimate of the error is simply the spread in the function $f(B-V)$.
\begin{equation}
{
\frac{\delta t}{t} = \frac{1}{n} \frac{\delta f}{f} \approx 2\frac{\delta f}{f}
}
\end{equation}

An estimate for $\delta f$ is the standard error of the residuals from the fit
to $f$, which we have derived using R, and which is $0.0795$ on
$182$ degrees of freedom. ($f$ itself as shown above, is of course known much
better because of the number of points involved.) For a K star, at $B-V = 1$, 
roughly the middle of our distribution, $f = 0.57$, so that
\begin{equation}
{
\frac{\delta t}{t} \approx 28\%
}
\end{equation}
The errors in $f$ are heteroscedastic, as can be seen from Fig.\,5, and on the
reasonable assumption that they scale with $f$, we can simply adopt this 
value of 28\% error in the ages for all G, K, \& early M stars.
This gives a representative number, but a uniform adoption of this error 
overestimates the age error for our stars\footnote{This can be traced to the
generous limits adopted in the present instance among the open clusters for 
inclusion as an I\,sequence star, resulting in considerable contamination from
incorrect rotation periods, g- and perhaps even C\,stars. This contamination
results in a large scatter in $f$, but $f$ itself is defined much better 
because of the large number of data points involved. Further work is needed in
open clusters to clarify this matter.}, 
and masks the underlying variations, which we elucidate below.

\subsection{Derivation of errors}

We begin with the representation
\begin{equation}
{
P = f(B-V).g(t)
}
\end{equation}
where $P, B-V$, and $t$ are the period, color and age of the star respectively, 
and $f$ and $g$ are the color and age dependencies, as before.
Taking logs and differentiating, we get
\begin{equation}
{
\frac{dP}{P} = \frac{df}{f} + \frac{dg}{g}
}
\end{equation}
Now, $g(t) = t^n$, so $dg/g = n dt/t + ln\,t\,dn$, where $n \approx 0.5$. Thus,
\begin{equation}
{
\frac{dP}{P} = \frac{df}{f} + n\frac{dt}{t} + ln\,t\,dn 
}
\end{equation}
Now, $f(B-V) = a x^b$, where $x=B-V-0.4$ and $a$ and $b$ are fitted constants
(with associated errors).
Differentiating, $df/f = da/a + b\,dx/x + ln\,x\,db$. Thus,
\begin{equation}
{
\frac{dP}{P} = \frac{da}{a} + b\frac{dx}{x} + ln\,x\,db + n\frac{dt}{t} + ln\,t\,dn
}
\end{equation}
Substituting, re-arranging, and adding the errors in quadrature under the usual
assumption of independence yields
\begin{equation}
{
(n \frac{\delta t}{t})^2 = (ln\,t\, \delta n)^2 + (\frac{\delta P}{P})^2 
                         + (\frac{\delta a}{a})^2 + (b\frac{\delta x}{x})^2 
                         + (ln\,x\,\delta b)^2
}
\end{equation}

The one term above that requires further attention is the period (P) term.
There are two contributions to the period error - the measurement error, and 
differential rotation, which can be added in quadrature:
 $(\delta P/P)^2 = (\delta P_{msrmnt}/P)^2 + (\delta P_{dffrtn}/P)^2$. 
The period determination itself is not usually a great contributor to the error,
but the differential rotation term could potentially be a deal-breaker.
Donahue et al. (1996) concluded that the dependence was a simple function of
the rotation period alone, and their results (see esp. Fig.\,3 there) suggest 
that the period range, $\Delta P = P_{max} - P_{min}$, 
can be represented simply by $ log\,(\Delta P) = -1.25 + 1.3\,log <P> $.
The long baseline of their dataset suggests that $\Delta P$ corresponds to 
$2 \sigma$, so that the ($1 \sigma$) period error is simply a quarter of this:
$log\,(\delta P_{dffrtn}) = -1.85 + 1.3\,log <P>$
so that,
\begin{equation}
{
(\frac{\delta P}{P})^2 = (\frac{\delta P_{msrmnt}}{P})^2 + (10^{-1.85}\,P^{0.3})^2 
}
\end{equation}

Substituting this in equation (10) gives
\begin{equation}
{
(n \frac{\delta t}{t})^2 =(ln\,t\, \delta n)^2 + (\frac{\delta P_{msrmnt}}{P})^2 
                         + (10^{-1.85}\,P^{0.3})^2   + (\frac{\delta a}{a})^2 
                         + (b\frac{\delta x}{x})^2 + (ln\,x\,\delta b)^2
}
\end{equation}

Putting in some of the numerical values will allow us to understand the 
dependencies of the errors.
From equation\,(2) - see appendix\,1 for the details - $n = 0.0519 \pm 0.007$. 
The error in the period determination for the Mt.\,Wilson stars is 
0.25\%-1\%, 0.5\%-2\%, 2\%-4\% for periods less than 20d, greater than 20d, 
or periods between 30d and 60d respectively (Donahue et al. 1996). A 1\% error 
seems to be a reasonable representation for the samples considered here.
For $\delta x = \delta (B-V) $, we adopt the value of 0.01 suggested by the
precision of the datasets considered below. This might need to be increased to 
0.02 for data acquired through CCD photometry (assuming independent errors of 
0.015 in each filter), but we note that this error could be considerably lower 
for data acquired through photoelectric photometry. 
From section\,3 and equation\,(1), 
$a = 0.7725 \pm 0.011 $ and $b = 0.601 \pm 0.024 $.
We input these values to get (in the same order as above)
\begin{equation}
{
(n\frac{\delta t}{t})^2 = (0.007\,ln\,t)^2 + (0.01)^2 + (0.014\,P^{0.3})^2 
                        + (\frac{0.011}{0.7725})^2 + (0.6\frac{0.01}{x})^2 
                        + (0.024\,ln\,x)^2
}
\end{equation}
or
\begin{equation}
{
(n\frac{\delta t}{t})^2 = 10^{-4}[  (0.7\,ln\,t)^2    + (1)^2 
                                 + (1.4\,P^{0.3})^2   + (1.424)^2
                                 + (\frac{0.6}{x})^2 + (2.4\,ln\,x)^2 
                                 ]
}
\end{equation}
or
\begin{equation}
{
(n\frac{\delta t}{t})^2 = 10^{-4}[ \frac{1}{2}(ln\,t)^2 + 1 + (1.4\,P^{0.3})^2 
                                   + 2 + (\frac{0.6}{x})^2 + (2.4\,ln\,x)^2
                                 ]
}
\end{equation}
Thus,
\begin{equation}
{
\frac{\delta t}{t} = 2\% \times \sqrt{3 + \frac{1}{2}(ln\,t)^2 + 2\,P^{0.6} 
                                          + (\frac{0.6}{x})^2  + (2.4\,ln\,x)^2
                                     }
}
\end{equation}
which shows that the age error is always greater than $\sim 11\%$ for conditions
similar to those assumed here. 
(Recall that $t$ is in Myr, $P$ is in days, and $x=B-V_0-0.4$.)
For 1\,Gyr-old stars of spectral types late\,F, early\,G, mid\,K and early\,M 
respectively, we get
\begin{equation}
{
\frac{\delta t}{t} = 2\% \times 
\cases{
\sqrt{26.9 + 6.4 + 66.5}  & when $B-V=0.5$  ($P= 7d$);\cr
\sqrt{26.9 + 8.9 + 16.9}   & when $B-V=0.65$ ($P=12d$);\cr
\sqrt{26.9 + 12.1 + 2.5}  & when $B-V=1.0$  ($P=20d$);\cr
\sqrt{26.9 + 15.4 + 0.35}  & when $B-V=1.5$  ($P=30d$).\cr
      }
}
\end{equation}
which shows the relative contributions of the period and color errors (second
and third terms, respectively), or,
\begin{equation}
{
\frac{\delta t}{t} = 
\cases{
20\%  & when $B-V=0.5$  ;\cr
15\%  & when $B-V=0.65$ ;\cr
13\%  & when $B-V=1.0$  ;\cr
13\%  & when $B-V=1.5$  .\cr
      }
}
\end{equation}
The behaviors of the function $f$ and of differential rotation are such that 
the color and period errors dominate for blue and red stars respectively to 
give a total error of $\sim$15\%. 
The errors calculated using equation (16) are the ones quoted for the 
gyrochronology ages in the remainder of this paper.
Setting the $P$ and $B-V$ errors equal leads to a trancendental equation which 
separates the color-period space into two regions, a blue one where color 
errors dominate, and a red one where the period errors (mostly differential
rotation) dominate. The separator is a steep function in color-period space, 
and is roughly at solar color.

How well these errors represent the true errors of this technique future work
will show. We simply note here that the very possibility of calculating the
errors distinguishes gyrochronology from other stellar chronometric methods.



\section{Comparison with the chromospheric clock} 

Before calculating gyro ages for stars where other ages are not available, it 
is necessary to consider whether these ages agree at least roughly with others 
that might be available.
We stress here that the ages derived through gyrochronology in this paper are 
{\it independent of other techniques}, except for the calibration using the Sun,
whose age is determined using radioactivity in meteorites (see prior section).
In particular, these ages are independent of chromospheric and isochrone ages,
except for the common solar calibration point\footnote{The calibration issue is
discussed later in this section.}.

Potentially, the best way to test these ages would be to derive ages for open
clusters where other ages are also available. This is not possible in this work 
because the open clusters have been used here to derive the mass dependence of 
gyrochronology, and this required a prior knowledge of their ages. 
Additional data will allow such a test in the future\footnote{In fact, two are underway using new data in M\,34 (James et al. 2007) and M\,35 (Meibom et al., in prep.)}.
Isochrone ages for main sequence field stars are not reliable enough to serve
as a test. (Section\,9 elaborates on this.) 
What is possible is a test against chromospheric ages for field stars.

Despite the obviously large errors associated with the method (see below), 
chromospheric ages have thus far been considered to be the best ones available 
at present for single field stars. 
Furthermore, there exists a substantial and uniform sample, the 
Mt.\,Wilson stars, for which the chromospheric emission is known very well 
(over decades), for which the chromospheric ages are believed to be relatively 
secure, and for which {\it measured} rotation periods are also 
available\footnote{There exists another sample of 19 Southern stars for which 
chromospheric emission (from Henry et al. 1996) and rotation periods are 
both available. These overlap with another sample of stars discussed here in 
section\,6 below, and the corresponding comparison is presented there.}. 
These facts allow us to compare the ages from the two (independent) 
techniques below.

\subsection{How is the chromospheric age of a star calculated?}

There has been considerable work on the determination of the rate of 
decay of chromospheric emission with age since the results of Skumanich (1972).
The two sources generally quoted for a relationship between chromospheric age 
and $R'_{HK}$ are Donahue (1998) and Soderblom et al. (1991).
Although we will end up using the former to calculate chromospheric age, 
it is necessary to discuss both to understand the relevant issues.
The key feature of both relationships is that once the measurement of stellar
chromospheric emission has been made (repeatedly or not),
it can immediately be converted into an age, without additional information. 
(Wright et al. 2004 have shown subsequently that stars previously considered to
be in Maunder Minima are in fact somewhat evolved, so caution is advisable
with respect to the basic properties of the star.)

\subsubsection{The Donahue (1998) relationship}

The relationship given in Donahue (1998) is:
\begin{equation}
{
log[t_{Chromospheric}] = 10.725 -1.334 R_5 + 0.4085 R_5^2 - 0.0522 R_5^3
}
\end{equation}
where $R_5 = 10^5 R'_{HK}$ and the age, $t$, is measured in Gyr.
This relationship is essentially identical to the one in Soderblom et al (1991) 
(discussed below) for ages greater than 1\,Gyr. The deviation between the two
relationships pertains to younger stars, including those in the Hyades, Coma,
Ursa Major, Pleiades, and NGC\,2264 open clusters, for which it claims a better
age calibration. This particular feature has prompted us to use it instead of 
the Soderblom et al. (1991) relationship.

However, it does have two serious limitations: 
Firstly, it does not provide errors on the ages so derived. (However, Donahue 
(1998) does list the discrepancies in chromospheric age for a number of wide 
binaries and triple systems. The mean discrepancy for the systems listed is 
0.85\,Gyr on a mean age of 1.85\,Gyr, which suggests a fractional age error of 
$\sim$46\%, in rough agreement with the errors quoted in Soderblom et al. 1991.)
Secondly,
there are no refereed publications that spell out the details of the derivation.
Nevertheless, it has been used by Wright et al. (2004) to derive ages for stars
in the sample being studied by the Marcy group for evidence of planets, 
and we follow suit.

\subsubsection{The Soderblom et al. (1991) relationship}

The relationship between chromospheric emission and age provided in
Soderblom et al. (1991) is
\begin{equation}
{
log[t_{Chromospheric}] = (-1.50 \pm 0.003) log R'_{HK} + (2.25 \pm 0.12)
}
\end{equation}
where the age, $t$, is again in Gyr. 
({\it Note especially that errors are provided.})
This expression, equation\,(3) from Soderblom et al. (1991), is based on
42 data points and three ``fundamental'' points, which are the Sun, the Hyades,
and the Ursa Major Group. This relationship passes through the data point for
the Sun, using the value of Log\,$R'_{HK}=-4.96$, quoted there, and a solar age
of 4.6\,Gyr. It is equivalent to $R'_{HK} \propto t^{-2/3}$, and 
the authors note that a case could be made for using slightly different 
relationships, including one where $R'_{HK} \propto t^{-3/4}$, equation (2) in 
their paper, depending on the choice of data points included. This relationship
has a standard deviation of 0.17\,dex, which corresponds to an error of 
$\sim$40\%.
In the absence of errors for the Donahue (1998) relationship above, we simply 
adopt this value of 0.17\,dex for the error in chromospheric ages calculated 
using that relationship also.

We note also that the Soderblom et al. (1991) relationship is in some ways the 
culmination of an extensive and self-consistent study by Soderblom and 
collaborators, and is explained in detail in a series of papers, including 
Duncan (1984), Duncan et al. (1984), Soderblom (1985), and  
Soderblom \& Clements (1987). The reader is referred to these for the 
technical details, and especially for the overall logic of the scheme.

One point about the calibration of the technique needs to be mentioned,
because it also relates to the calibration of gyrochronology.
The ages against which the above relationship is calculated are derived using
isochrone fits to visual binary stars, and to the ``fundamental'' points, which 
are again based on isochrone fits. 
The entire isochrone technique itself is calibrated by ensuring
that the appropriate solar model matches the solar parameters, usually the
radius and the luminosity, at solar age. Thus, this technique is also ultimately
calibrated on the Sun.

\subsection{Comparison between chromospheric and gyro ages for the Mt.\,Wilson stars}

We use the data compilation published in Baliunas et al. (1996) and
Noyes et al. (1984) for the Mt.\,Wilson stars and calculate the chromospheric 
ages using the formula in Donahue (1998).
The chromospheric ages for the Mt.\,Wilson stars, calculated using the above
formula, are listed in Table\,3. For obvious reasons, stars with calculated 
periods have been excised, and only stars with measured periods 
(71 in number) have been retained for this comparison\footnote{We have also
had to eliminate HD124570, which, although not considered evolved in the Mt.\,
Wilson datasets, is now known to be so (e.g. Cowley, 1976; SAB thanks Brian
Skiff for researching this star).}.

For the same stars, we can calculate ages via gyrochronology using 
equation\,(3) from Section\,3 above.
These ages are calculated and listed in Table\,3. 
The errors on these ages, calculated using equation\,(16) from section\,4,
are also listed in the table. 

The gyro ages are plotted against the chromospheric ages for the same stars
in Fig.\,7, with small green and large red symbols marking the young (Y) and 
old (O) Mt.\,Wilson stars, as classified by Vaughan (1980). 
Note that both techniques segregate the Y and O stars. The demarcation is 
sharper in chromospheric age, as it ought to be, since this is the criterion 
chosen to classify the stars as young or old.
The figure shows that, apart from a slight tendency towards shorter gyro ages
(discussed further below), there is general agreement between the 
chromospheric- and gyro ages for this sample.
Note that except for a few stars discussed below, there are no stars with 
widely discrepant ages, unlike the corresponding comparison with isochrone ages,
where discrepancies are routine (c.f. Fig.\,2 in Barnes 2001). 

Note especially that the gyro ages are well-behaved for every single star here,
ranging from just under 100\,Myr to just over 10\,Gyr. In contrast, there
are three stars whose chromospheric ages are almost certainly incorrect
(see table\,4).
The two stars HD82443 and HD129333 have chromospheric ages of 0.67\,Myr 
and 0.002\,Myr respectively. These stars are undoubtedly young but the 
interpretation of these numbers eludes us. The corresponding gyro ages 
for these stars are $164 \pm 18$\,Myr and $73 \pm 9$\,Myr, 
which suggest that they are essentially on the Zero Age Main Sequence (ZAMS).
Also, for one star, HD95735, the chromospheric age is 20\,Gyr, greater than 
that of the universe (dotted lines). This cannot be correct. The gyro age for
this star is $3.2 \pm 0.5$\,Gyr, definitely younger than the Sun.
At least in this restricted sense and for specific stars, the gyro ages are 
better defined than chromospheric ages.

Fig.\,8 elaborates on the difference between the chromospheric- and gyro ages. 
The solid line again denotes equality, while the dashed line, 
at $t_{gyro}=0.74 \times t_{chrom}$, bisects the data points. This shows that
the gyro ages are roughly 25\% lower than the chromospheric ages overall. 
The nature of the disagreement can be probed by segregating the stars by color.
Thus, stars bluer than $B-V=0.6$ and redder than $B-V=0.8$ are plotted using
blue crosses and red asterisks respectively, while those with intermediate
colors are plotted using green squares.
This exercise shows that there is good agreement for stars redward of 
$B-V=0.6$ and that the above discrepancy pertains only to the blue stars.

\subsection{Discussion of the disagreement between the techniques}

This disagreement can be probed further than merely stating that the errors in
the chromospheric ages are greater than those of the gyro ages. It must
originate in either the gyro ages for blue stars being systematically shorter
or the chromospheric ages for these being systematically longer, or both.
The discussion below, and the results of testing binaries, performed in 
Section\,8, suggest that the chromospheric ages are the more problematical.

With respect to the gyro ages, one defect is that
the open cluster sample used to define the mass dependence of rotation
does not contain stars with $B-V$ colors blueward of 0.5 because it is not yet 
possible to distinguish between very blue C- and I-type stars. This means that
$f(B-V)$ is an extrapolation for stars with $ 0.4 < B-V < 0.5$. The fitting 
function $f(B-V)$, blueward of $B-V=0.6$, appears to be somewhat elevated with 
respect to the data points displayed in Fig.\,5. This would tend to lower the 
gyro ages. If $f(B-V)$ were lowered in this region by $\sim$20\%, the gyro ages 
would be raised by a factor of $\sim$1.5, which is doable considering the 
main sequence lifetime of the F\,stars, but a reduction of $\sim$30\% would
double the gyro ages, and might run afoul of standard stellar evolution,
because the main sequence lifetime of a late-F star is $\sim$5\,Gyr.

The chromospheric ages are not blamefree in this regard either, and it is
almost certain that they have been overestimated for F\,stars\footnote{The embedded mass dependence in the chromospheric ages can be traced to Noyes et al. (1984), where the mass dependence of chromospheric emission was based on the Rossby Number, and theoretical estimates of the variation of convective turnover timescale with stellar mass. The residual mass dependence could be removed eventually with the availability of larger samples of stars, especially those in open clusters.}. 
Four F\,stars have chromospheric ages in excess of 7\,Gyr and one in excess of 
6\,Gyr. These values exceed the main sequence lifetime of a late-F star, which 
is 5\,Gyr.
In comparison, all of these five F\,stars have shorter gyro ages, with the 
oldest of them assigned a gyro age of 2.3\,Gyr.
These stars are also listed in Table\,4. These stars are located at higher
chromospheric ages than 5\,Gyr, marked in Fig.\,8 with thin dashed blue lines.
Thus, while the gyro ages have possibly been slightly underestimated for 
F\,stars, it is almost certain that the corresponding chromospheric ages have 
been overestimated.

\subsection{Additional issues with chromospheric ages}

The derivation of a chromospheric age for a star is complicated by the
natural variability of chromospheric emission with stellar rotational phase
and stellar cycle (e.g. Wilson, 1963).
In fact, rotation-related variations in chromospheric emission are the 
preferred way of deriving rotation periods for old stars.
Binarity or other effects could result in additional variability.
These variations make it necessary for repeated measurement on a suitable
timescale of the chromospheric emission from a star to ensure that the 
measured average is a good representation of the chromospheric emission at 
that age for the star.
Therefore, it is unlikely that one can make a single measurement of 
chromospheric emission and derive a good age for a star.

Some of the issues with chromospheric ages are illustrated by the recent work
of Giampapa et al. (2006) on the chromospheric properties of the sun-like
stars in the open cluster M\,67, averaged over several seasons of observing. 
This cluster is known to be $\sim$4\,Gyr old (e.g. VandenBerg \& Stetson, 2004).
There is no evidence that the stars in this cluster are not coeval.

Correspondingly, Giampapa et al. (2006) derive mean and median ages in the 
range 3.8\,Gyr to 4.3\,Gyr.
What is surprising is that the chromospheric ages for individual stars
range from under 1\,Gyr to 7.5\,Gyr (see Fig.\,13 in their paper).
Admittedly, the vast majority of the stars have chromospheric ages between
2\,Gyr and 6\,Gyr, but this range is not small either. 
This result seems to cast doubts on the precision in ages for single stars 
obtainable even in principle with chromospheric emission because measuring
the age for M\,67 using a random cluster member could result in such large
age variability.


\clearpage

\begin{deluxetable}{llcrrrrr}
\tabletypesize{\scriptsize}
\tablecaption{Gyrochronology ages and errors for the Mt.\,Wilson stars. \label{tbl-3}}
\tablewidth{0pt}
\tablehead{
\colhead{HD} & \colhead{$B-V$}   &\colhead{$P_{rot}$(d)\tablenotemark{A}} & \colhead{$-log<R'_{HK}>$} & \colhead{$t_{chrom}$/Myr } & \colhead{$t_{iso}$/Myr\tablenotemark{B}} & \colhead{$t_{gyro}$/Myr}   & \colhead{$\delta t_{gyro}$/Myr} 
}
\startdata
Sun	&0.642 &26.09\tablenotemark{C} &4.901 &3895	   &...... &4566 &770\\
1835   	&0.66  &7.78 	 &4.443 &601       &$<$1760  &408     &54\\
2454 	&0.43  &3 	 &4.792 &2609        &......   &790     &350\\
3229 	&0.44  &2 	 &4.583 &1251        &......   &260     &91\\
3651 	&0.85  &44 	 &4.991 &5411       &$>$11800   &6100    &990\\
4628 	&0.88  &38.5 	 &4.852 &3250       &$>$6840   &4370    &680\\
6920 	&0.60  &13.1 	 &4.793 &2618        &......   &1510    &240\\
10476 	&0.84  &35.2 	 &4.912 &4056       &$>$8840   &4070    &630\\
10700 	&0.72  &34 	 &4.958 &4802      &$>$12120   &5500    &910\\
10780 	&0.81  &23 	 &4.681 &1764         &10120   &1945    &280\\
16160 	&0.98  &48.0 	 &4.958 &4802           &540   &5370    &850\\
16673 	&0.52  &7 	 &4.664 &1664        &......   &820     &150\\
17925 	&0.87  &6.76 	 &4.311 &74         &$<$1200   &157     &16\\
18256 	&0.43  &3 	 &4.722 &2033        &......   &790     &350\\
20630 	&0.68  &9.24 	 &4.420 &489         &$<$2760   &522     &69\\
22049 	&0.88  &11.68 	 &4.455 &659 	   &$<$600   &439     &52\\
25998 	&0.46  &2.6 	 &4.401 &398         &......   &270     &70\\
26913 	&0.70  &7.15 	 &4.391 &352         &......   &294     &36\\
26965 	&0.82  &43 	 &4.872 &3499       &$>$9280   &6320    &1030\\
30495 	&0.63  &7.6\tablenotemark{D} &4.511 &923 &6080   &450     &62\\
35296 	&0.53  &3.56 	 &4.378 &294 	   &......   &202     &33\\
37394 	&0.84  &11 	 &4.454 &654 	  &$<$1360   &432     &52\\
39587 	&0.59  &5.36 	 &4.426 &518 	     &4320   &286     &41\\
45067 	&0.56  &8 	 &5.094 &7733 	     &5120   &760     &120\\
72905 	&0.62  &4.69 	 &4.375 &281 	   &......   &187     &24\\
75332 	&0.49  &4 	 &4.464 &703 	     &1880   &387     &79\\
76151 	&0.67  &15 	 &4.659 &1635 	     &1320   &1380    &200\\
78366 	&0.60  &9.67 	 &4.608 &1370        &$<$680   &840     &130\\
81809 	&0.64  &40.2 	 &4.921 &4193 	   &......   &10600\tablenotemark{E} &1900\\
82443 	&0.77  &6 	 &4.211 &0.7         &......   &164     &18\\
89744 	&0.54  &9 	 &5.120 &8421 	     &1880   &1110    &190\\
95735 	&1.51  &53 	 &5.451 &20028 	   &......   &3070    &460\\
97334 	&0.61  &8 	 &4.422 &499 	  &$<$2920   &551     &80\\
100180 	&0.57  &14 	 &4.922 &4209 	     &3800   &2070    &350\\
101501 	&0.72  &16.68 	 &4.546 &1082      &$>$11320   &1400    &200\\
106516 	&0.46  &6.91 	 &4.651 &1591 	   &......   &1770    &480\\
107213 	&0.50  &9 	 &5.103 &7966 	     &2040   &1630    &330\\
114378 	&0.45  &3.02 	 &4.530 &1010 	   &......   &445     &130\\
114710 	&0.57  &12.35 	 &4.745 &2205       &$<$1120   &1630    &270\\
115043 	&0.60  &6 	 &4.428 &528 	   &......   &335     &47\\
115383 	&0.58  &3.33 	 &4.443 &601         &$<$760   &122     &17\\
115404 	&0.93  &18.47 	 &4.480 &779         &......   &950     &120\\
115617 	&0.71  &29 	 &5.001 &5609 	     &8960   &4200    &680\\
120136 	&0.48  &4 	 &4.731 &2098 	     &1640   &443     &97\\
129333 	&0.61  &2.80 	 &4.152 &0.002      &$<$1440   &73 	    &9\\
\hline 
131156A &0.76  &6.31\tablenotemark{F}	 &4.363 &232	   &$<$760   &187	&21\\
131156B &1.17  &11.94\tablenotemark{G} &4.424 &508 	 &$>$12600   &265     &28\\
\hline 
141004	&0.60  &25.8	 &5.004 &5669          &6320   &5570    &990\\
143761 	&0.60  &17 	 &5.039 &6413 	     &9720   &2490    &410\\
149661 	&0.82  &21.07 	 &4.583 &1251       &$<$4160   &1600    &220\\
152391 	&0.76  &11.43 	 &4.448 &625          &720     &587     &75\\
154417 	&0.57  &7.78 	 &4.533 &1023         &4200   &670     &105\\
\hline 
155885	&0.86  &21.11\tablenotemark{H} &4.559 &1141	   &......   &1440    &200\\
155886 	&0.86  &20.69\tablenotemark{I} &4.570 &1191 	   &......   &1390    &190\\
156026 	&1.16  &18.0\tablenotemark{J}  &4.622 &1439 	   &$<$480   &593     &71\\
\hline 
160346	&0.96  &36.4	 &4.795 &2637	   &......   &3280    &490\\
165341A\tablenotemark{K}  &0.86 &20  &4.548 &1091 	   &......   &1300    &180\\
166620 	&0.87  &42.4 	 &4.955 &4750      &$>$11200   &5400    &860\\
178428 	&0.70  &22 	 &5.048 &6616        &......   &2560    &390\\
185144 	&0.80  &27 	 &4.832 &3019 	   &......   &2730    &410\\
187691 	&0.55  &10 	 &5.026 &6128 	     &3200   &1250    &210\\
190007 	&1.17  &28.95 	 &4.692 &1832       &$<$1760   &1460    &200\\
190406 	&0.61  &13.94 	 &4.797 &2657          &3160   &1610    &250\\
194012 	&0.51  &7 	 &4.720 &2019 	   &......   &900     &170\\
\hline 
201091	&1.18  &35.37	 &4.764 &2359        &$<$440   &2120    &300\\
201092 	&1.37  &37.84\tablenotemark{L} &4.891 &3753   &$<$680   &1870    &260\\
\hline 
206860	&0.59  &4.86	 &4.416 &470         &$<$880   &237	    &33\\
207978 	&0.42  &3 	 &4.890 &3740        &......   &1270    &810\\
212754 	&0.52  &12 	 &5.073 &7207 	   &......   &2300    &440\\
219834B\tablenotemark{M}  &0.91 &43  &4.944 &4563 &$>$13200   &5040   &800\\
224930 	&0.67  &33   	 &4.875 &3538 	   &......   &6330    &1080\\
\enddata


\tablenotetext{A}{Only {\it measured} periods for unevolved stars are listed.
They are taken, in order of priority, from Donahue, Saar \& Baliunas (1996), 
Baliunas et al. (1983), and Baliunas, Sokoloff \& Soon (1996). The first of
these lists the average rotation period of several seasonal periods (and the 
differential rotation), hence the priority assigned to this paper, the second 
a single best period determined from an intensive chromospheric monitoring
program in 1980-81 (with the error of that single determination), and the third
a mean rotation period (to lower precision than the previous two publications)
based on the entire extant intensive sampling database.}
\tablenotetext{B}{The isochrone ages listed in this and subsequent tables are taken from Takeda et al. (2007).}
\tablenotetext{C}{The mean solar period of 26.09d, taken from Donahue et al.
(1996), represents the average of 8 determinations, and is presumably 
representative of the mean latitude of sunspot persistence, while the $\sim$25d
period usually listed is the mean equatorial rotation period.}
\tablenotetext{D}{Baliunas et al. (1996) list a significantly different period
of 11d.}
\tablenotetext{E}{The gyro age should be treated with caution because this star
is a spectroscopic binary (Pourbaix, 2000).}
\tablenotetext{F}{Period is from Donahue et al. (1996). Baliunas et al. (1996) simply list a period of 6d.}
\tablenotetext{G}{Period is from Donahue et al. (1996). Baliunas et al. (1996) list a period of 11d.}
\tablenotetext{H}{This period is from Donahue et al. (1996). Baliunas et al. (1983) list a very similar period of 22.9$\pm$0.5d.}
\tablenotetext{I}{This period is from Donahue et al. (1996). Baliunas et al. (1983) list a very similar period of 20.3$\pm$0.4d.}
\tablenotetext{J}{This period is from Baliunas et al. (1983). Baliunas et al. (1996) list a period of 21d for all 3 components HD155885, HD155886, \& HD156026.}
\tablenotetext{K}{The second component, HD165341B, of this binary is also in the Mt.\,Wilson sample (e.g. Baliunas et al. 1996), but the period of 34d is one calculated from chromospheric emission.}
\tablenotetext{L}{The periods listed for HD201091 and HD201092 are from Donahue et al. (1996). Baliunas et al. (1996) list similar periods of 35 and 38d respectively, while Baliunas et al. (1983) list the somewhat discrepant periods of 37.9$\pm$1.0 and 48d respectively.}
\tablenotetext{M}{The other component in this system, HD219834A, is also in the Mt.\,Wilson dataset, but it seems to be evolved, and so is excluded here.}

\end{deluxetable}

\clearpage

\begin{deluxetable}{llrrl}
\tabletypesize{\scriptsize}
\tablecaption{Stars with suspect chromospheric ages. \label{tbl-4}}
\tablewidth{0pt}
\tablehead{
\colhead{Star} & \colhead{$B-V$}   & \colhead{$Age_{chromo}$} & \colhead{$Age_{gyro}$} & \colhead{Comment}   
}
\startdata
HD45067  & 0.56   & 7.73\,Gyr  & 0.76$\pm$0.1\,Gyr & Chromo age $>$ lifetime\\ 
HD82443  & 0.77   & 0.7\,Myr  &  164$\pm$18\,Myr & Chromo age too small?\\
HD89744  & 0.54   & 8.42\,Gyr  & 1.11$\pm$0.19\,Gyr & Chromo age $>$ lifetime\\
HD95735  & 1.51   & 20\,Gyr    & 3.1$\pm$0.46\,Gyr & Chromo age $>$ Age of universe\\
HD107213 & 0.50   & 7.97\,Gyr  & 1.63$\pm$0.33\,Gyr & Chromo age $>$ lifetime\\
HD129333 & 0.61   & 0.002\,Myr &  73$\pm$9\,Myr &  Chromo age too small?\\
HD187691 & 0.55   & 6.13\,Gyr  & 1.25$\pm$0.21\,Gyr &  Chromo age $>$ lifetime\\
HD212754 & 0.52   & 7.21\,Gyr  & 2.3$\pm$0.44\,Gyr &  Chromo age $>$ lifetime\\
\enddata




\end{deluxetable}

\section{Ages for young field stars from the Vienna-KPNO (Strassmeier et al. 2000) survey}

Another group of stars amenable to the calculation of ages via gyrochronology
is the field star sample of Strassmeier et al. (2000). Unlike the older 
Mt.\,Wilson star sample, this group contains some stars for which gyro ages 
are not yet appropriate, and here we demonstrate how to identify and excise 
these stars and calculate ages for the rest.

The full Strassmeier et al. (2000) sample consists of 1058 Hipparcos stars with 
various measured parameters, including chromospheric emission. 
Of these 1058 stars, 140 have measured rotation periods, and of these, 
we are interested here only in stars on the main sequence. Using the Luminosity 
classes supplied by Strassmeier et al., we have simply selected the dwarf stars,
and excised the others. 
This leaves us with 101 dwarf stars with measured rotation periods.

These 101 stars with measured periods are plotted in a color-period diagram in
Fig.\,9. We superimpose an I sequence curve corresponding to 100\,Myr, and 
assume that the 16 stars below this curve are C-\,or g\,stars, while those
above are I\,sequence stars similar to those in the Mt.\,Wilson sample. 
In the scenario from Barnes (2003a), the stars below are either on the 
C\,sequence appropriate to their age, or in the transition, g, between the 
C-\,and I\,sequences.
This cut is undoubtedly conservative, since there are open clusters younger
than 100\,Myr known to possess I sequences, but we prefer to lose a few stars
rather than risk over-extending the technique.
This leaves us with 85 potential I\,sequence stars amenable to gyrochronology. 

These 85 I\,sequence stars are again plotted in Fig.\,10, where now we have
superimposed  isochrones corresponding to ages of
100-, 200-, \& 450\,Myr, and 1-, 2-, \& 4.5\,Gyr. 
We see that the Strassmeier et al. (2000) main sequence sample with measured 
periods consists mainly of stars younger than 1\,Gyr, and all but 4 younger than
2\,Gyr. In fact the median age for the sample is 365\,Myr, in keeping with the 
selection of this sample for activity.

Gyro ages are calculated as above for each star, and listed in Table\,5, along
with their basic measured properties.
Almost all of these stars are redder than the Sun.
For such stars, as shown in the previous section, there is very good agreement
between gyro- and chromospheric ages, and consequently, some confidence can be
attached to the calculated ages. 
We have also calculated the errors on these ages, using equation\,(16) from
section\,4, and listed these in the final column of the table.

At present, no good test of these ages is possible. 
Although the chromospheric emission has been measured, the measurements have
been made with a small telescope, and are not long-term averages, so that
the values quoted cannot be treated with the confidence associated, for
instance, with the Mt.\,Wilson measurements, and there is considerable scatter,
as the few repeat measurements demonstrate. 
Furthermore, no photospheric correction has been performed, so they 
are on a different scale, and the relationships of Soderblom et al. (1991) and 
Donahue (1998) do not apply. 
However, it is possible to plot the $R_{HK}$ values provided against the gyro 
ages calculated above to make sure that gross errors are absent, and we perform 
this exercise in Fig.\,11.
The figure demonstrates that, as expected, the chromospheric activity declines 
steadily with stellar age, and thus, that the gyro ages are reasonable. 


\clearpage

\begin{deluxetable}{lrcrrrr}
\tabletypesize{\scriptsize}
\tablecaption{Gyrochronology ages and errors for the Vienna-KPNO survey (Strassmeier et al. 2000) stars. \label{tbl-5}}
\tablewidth{0pt}
\tablehead{
\colhead{HD} & \colhead{$B-V$}   &$P_{rot}$(d) & \colhead{$R_{HK}$} & \colhead{$t_{iso}$/Myr} & \colhead{$t_{gyro}$/Myr}   & \colhead{$\delta t_{gyro}$/Myr} 
}
\startdata
HD691      &0.76 &6.105  &7.2E-5 &$<$1040  &175     &20\\
HD5996     &0.76 &12.165 &6.0E-5 &......   &662     &86\\
HD6963     &0.73 &20.27  &4.9E-5 &2240     &1960    &290\\
HD7661     &0.75 &7.85   &7.1E-5 &......   &294     &35\\
HD8997a    &0.97 &10.49  &3.5E-5 &......   &292     &32\\
HD8997b    &0.97 &10.49  &1.8E-5 &......   &292     &32\\
HD9902b    &0.65 &7.41   &8.9E-5 &......   &389     &52\\
HD10008    &0.80 &7.15   &6.1E-5 &......   &211     &23\\
HD12786    &0.83 &15.78  &5.6E-5 &......   &890     &120\\
HD13382    &0.68 &8.98   &6.0E-5 &......   &494     &65\\
HD13507    &0.67 &7.60   &6.6E-5 &$<$1320  &373     &48\\
HD13531    &0.70 &7.52   &7.0E-5 &$<$3520  &324     &40\\
HD13579A   &0.92 &6.79   &2.8E-5 &$>$8840  &141     &14\\
HD16287    &0.94 &11.784 &5.3E-5 &$<$2360  &390     &45\\
HD17382    &0.82 &$>$50  &6.1E-5 &......   &$>$8450\tablenotemark{A} &$>$1400\\
HD18632    &0.93 &10.055 &5.3E-5 &$>$7800  &293     &33\\
HD18955a   &0.86 &8.05   &6.1E-5 &......   &225     &25\\
HD19668    &0.81 &5.41   &4.1E-5 &......   &120     &12\\
HD19902    &0.73 &$>$50  &5.4E-5 &......   &$>$11200 &$>$2000\\
HD20678    &0.73 &5.95   &6.6E-5 &......   &185     &21\\
HD27149a   &0.68 &8.968  &5.7E-5 &......   &492     &65\\
HD27149b   &0.68 &8.968  &5.1E-5 &......   &492     &65\\
HD28495    &0.76 &7.604  &9.3E-5 &......   &268     &31\\
HD31000    &0.75 &7.878  &8.2E-5 &......   &296     &35\\
HD53157    &0.81 &10.88  &6.2E-5 &......   &460     &56\\
HD59747    &0.86 &8.03   &6.5E-5 &$<$920   &224     &24\\
HD73322    &0.91 &16.41  &5.1E-5 &......   &788     &100\\
HD75935    &0.77 &8.19   &6.3E-5 &......   &299     &35\\
HD77825    &0.96 &8.64   &5.2E-5 &......   &205     &22\\
HD79969    &0.99 &43.4   &3.6E-5 &......   &4340    &670\\
HD82443    &0.78 &5.409  &9.4E-5 &......   &130     &14\\
HD83983    &0.88 &10.92  &3.9E-5 &......   &386     &45\\
HD87424    &0.89 &10.74  &5.3E-5 &$<$1720  &365     &42\\
HD88638    &0.77 &4.935  &8.4E-5 &......   &113     &12\\
HD92945    &0.87 &13.47  &7.5E-5 &$<$1280  &592     &73\\
HD93811    &0.94 &8.47   &5.0E-5 &......   &206     &22\\
HD94765    &0.92 &11.43  &4.6E-5 &$<$1480  &384     &44\\
HD95188    &0.76 &7.019  &7.1E-5 &$<$960   &230     &26\\
HD95724    &0.94 &11.53  &5.2E-5 &......   &374     &43\\
HD95743    &0.97 &10.33  &4.0E-5 &......   &284     &31\\
HD101206   &0.98 &10.84  &4.1E-5 &......   &305     &34\\
HD103720   &0.95 &17.16  &4.2E-5 &......   &787     &99\\
HD105963A  &0.88 &7.44   &6.6E-5 &......   &184     &19\\
HD105963B  &0.88 &7.44   &6.2E-5 &......   &184     &19\\
HD109011a  &0.94 &8.31   &5.8E-5 &......   &199     &21\\
HD109647   &0.95 &8.73   &5.3E-5 &......   &214     &23\\
HD110463   &0.96 &11.75  &4.5E-5 &......   &371     &42\\
HD111813   &0.89 &7.74   &6.5E-5 &......   &194     &21\\
HD113449   &0.85 &6.47   &6.9E-5 &......   &152     &16\\
HD125874   &0.88 &7.52   &6.1E-5 &......   &188     &20\\
HD128311   &0.97 &11.54  &4.5E-5 &$<$960   &351     &40\\
HD130307   &0.89 &21.79  &3.9E-5 &1520     &1425    &190\\
HD139194   &0.87 &9.37   &3.8E-5 &......   &294     &33\\
HD139837   &0.73 &6.98   &8.0E-5 &......   &251     &30\\
HD141272   &0.80 &14.045 &6.7E-5 &......   &773     &100\\
HD141919   &0.88 &13.62  &4.4E-5 &......   &590     &72\\
HD142680   &0.97 &33.52  &2.5E-5 &......   &2740    &400\\
HD144872   &0.96 &26.02  &2.7E-5 &......   &1720    &240\\
HD150511   &0.88 &10.58  &5.0E-5 &......   &363     &42\\
HD153525   &1.00 &15.39  &1.7E-5 &......   &577     &69\\
HD153557   &0.98 &7.22   &3.7E-5 &......   &140     &14\\
HD161284   &0.93 &18.31  &4.2E-5 &......   &930     &120\\
HD168603   &0.77 &4.825  &6.4E-5 &......   &108     &11\\
HD173950   &0.83 &10.973 &5.2E-5 &......   &442     &53\\
HD180161   &0.80 &5.49   &2.7E-5 &......   &127     &13\\
HD180263   &0.91 &14.16  &3.7E-5 &......   &593     &72\\
HD189733   &0.93 &12.039 &4.6E-5 &......   &415     &48\\
HD192263   &0.94 &23.98  &4.1E-5 &2560     &1530    &210\\
HD198425   &0.94 &22.64  &4.3E-5 &......   &1370    &185\\
HD200560   &0.97 &10.526 &5.2E-5 &......   &294     &32\\
HD202605   &0.74 &13.78  &5.3E-5 &......   &900     &120\\
HD203030   &0.75 &6.664  &7.1E-5 &......   &215     &25\\
HD209779   &0.67 &10.29  &6.2E-5 &10000    &670     &92\\
HD210667   &0.81 &9.083  &5.3E-5 &$<$4200  &325     &38\\
HD214615AB &0.76 &6.20   &7.5E-5 &......   &181     &20\\
HD214683   &0.94 &18.05  &4.2E-5 &......   &886     &110\\
HD220182   &0.80 &7.489  &6.4E-5 &......   &230     &26\\
HD221851   &0.85 &12.525 &3.8E-5 &......   &541     &66\\
HD258857   &0.91 &19.98  &3.9E-5 &......   &1150    &150\\
HIP36357   &0.92 &11.63  &4.7E-5 &......   &397     &46\\
HIP43422   &0.75 &11.14  &3.3E-4 &......   &578     &74\\
HIP69410   &0.96 &9.52   &5.0E-5 &......   &248     &27\\
HIP70836   &0.94 &21.84  &3.2E-5 &......   &1280    &170\\
HIP77210ab &0.83 &13.83  &6.3E-5 &......   &690     &87\\
HIP82042   &0.96 &13.65  &3.2E-5 &......   &496     &59\\
\enddata


\tablenotetext{A}{The gyro age should be treated with caution because this star
is a spectroscopic binary (Latham et al. 2002).}


\end{deluxetable}

\section{Ages for young- and intermediate-age field stars from Pizzolato et al. (2003) with X-ray measurements}

There exists another comparably large group of main sequence field stars for 
which rotation periods and other relevant information are available. This group 
has been assembled by Pizzolato et al. (2003) in connection with a study of 
X-ray activity. There are 110 stars in this group. We remove 2 of these, 
HD82885 and HD136202, suspected to be evolved, leaving 108 stars. 
51 of these 108 stars are in also in the Mt.\,Wilson sample, but the 
remainder do not overlap with the Strassmeier et al. (2000) stars either,
and hence warrant attention. Furthermore, these stars also have measured X-ray
fluxes, listed conveniently in Pizzolato et al. (2003), 
which allow a crude comparison with the gyro ages we derive below.

The Pizzolato et al. (2003) stars must follow the same rotational patterns as 
the open cluster, Mt.\,Wilson, and Strassmeier et al. (2000) stars. 
We can use the same condition that we used with the Strassmeier et al. (2003) 
stars to excise the C/g stars from the sample.
We plot the color-period diagram for this sample in Fig.\,12.
Plotting a 100\,Myr isochrone as before, we excise all stars below it,
since these are either C- or g\,type stars, or only ambiguously I\,type.
Note that of the excised stars, the ones with periods below 1 day are
almost certainly C\,sequence stars. 
We also excise GL551 ($B-V=1.90$, $P=42d$) because it is fully convective
and therefore unable to sustain an interface dynamo.
(Note also that apart from this one object, there are no slow rotators redward 
of $B-V=1.55$.  This is consistent with the prediction by Barnes (2003a) for 
the terminus of the I sequence at the point of full convection.)
This leaves us with 79 stars that are potentially on the I sequence in this 
sample.

These 79 stars are suitable for gyrochronology. 
We calculate the gyro ages as before, and list them, their errors, and other 
relevant information for these stars in Table\,6.
Fig.\,13 displays the color-period diagram for these 79 stars, with isochrones 
at 100\,Myr, 200\,Myr, 450\,Myr, 1\,Gyr, 2\,Gyr, 4.5\,Gyr, \& 10\,Gyr.
Fig.\,13 shows that this sample spans a substantial range of ages, 
from 100\,Myr to 6\,Gyr (all but 4 of them), 
although most of them are younger than the Sun, and the median age for the
sample is 1.2\,Gyr.

These results are reasonably consistent with Sandage et al. (2003) who suggest a
(classical isochrone) age for the oldest stars in the local Galactic disk of
7.4 to 7.9\,Gyr ($\pm 0.7$\,Gyr) depending on whether or not the stellar
models allow for diffusion.
All the Pizzolato et al. (2003) stars except for HD81809, 
which is known to be a spectroscopic binary (Pourbaix, 2000),
have calculated gyro ages shortward of this age.

The available X-ray data for these same stars also suggest that the gyro
ages are reasonable. In Fig.\,14, we plot the X-ray emission from these stars
against their gyro ages. We see that the X-ray emission declines steadily,
as expected, and in fact, there are no widely discrepant data points. 

Finally, we note that in addition to the 51 stars in this group that are common
to the Mt.\,Wilson sample, 19 are present in the chromospheric emission survey
of Southern stars by Henry et al. (1996), where their $R'_{HK}$ values are
published. These are on the same system as the Mt.\,Wilson data. Thus, it is
possible to compute their chromospheric ages, and compare them with the ages
from gyrochronology. This comparison is shown in Fig.\,15. 
The stars can be seen to scatter around the line of equality, and in fact,
the agreement between the gyro and chromospheric ages for all but two of them
is within a factor of two (see Fig.\,15).


\clearpage

\begin{deluxetable}{llcrrrr}
\tabletypesize{\scriptsize}
\tablecaption{Gyrochronology ages and errors for the Pizzolato et al. (2003) stars. \label{tbl-6}}
\tablewidth{0pt}
\tablehead{
\colhead{Star} & \colhead{$B-V$}   &$P_{rot}$(d) &$log(L_x/L_{bol})$ & \colhead{$t_{iso}$/Myr}   & \colhead{$t_{gyro}$/Myr} & \colhead{$\delta t_{gyro}$/Myr}
}
\startdata
Sun          &0.66       &25.38     &-6.23	&........      & 3980   & 650\\
GL338B      &1.42       &10.17     &-4.90	&........      & 140    & 14\\
GL380       &1.36       &11.67     &-5.04	&........      & 196    & 20\\
GL673       &1.36       &11.94     &-4.96	&........      & 205    & 21\\
GL685       &1.45       &18.60     &-4.92	&........      & 435    & 50\\
HD1835      &0.66       & 7.70     &-4.67	&$<$1760      &  400    & 53\\
HD3651      &0.85       &48.00     &-5.70	&$>$11800     & 7200   & 1200\\
HD4628      &0.88       &38.00     &-6.01	&$>$6840      &  4260   & 660\\
HD10360     &0.88       &30.00     &-5.97	&$<$600      &   2700   & 400\\
HD10361     &0.86       &39.00     &-5.94	&$<$520      &   4710   & 740\\
HD10476     &0.84       &35.20     &-6.59	&$>$8840      &  4070   & 630\\
HD10700     &0.72       &34.50     &-6.21	&$>$12120      & 5660   & 930\\
HD11507     &1.43       &15.80     &-4.76	&........      & 324    & 36\\
HD13445     &0.82       &30.00     &-5.56	&$>$8480      &  3160   & 480\\
HD14802     &0.60       & 9.00     &-4.50	&........      & 732    & 110\\
HD16160     &0.97       &45.00     &-5.71	&   540      &   4840   & 760\\
HD16673     &0.52       & 7.40     &-5.03	&........      & 910    & 170\\
HD17051     &0.56       & 7.90     &-5.02	&  2720      &   740    & 120\\
HD17925     &0.87       & 6.60     &-4.51	&$<$1200      &  150    & 15\\
HD20630     &0.68       & 9.40     &-4.62	&$<$2760      &  540    & 72\\
HD22049     &0.88       &11.30     &-4.92	&$<$600      &   412    & 48\\
HD25998     &0.52       & 3.00     &-4.40	&........      & 159    & 27\\
HD26913     &0.70       & 7.20     &-4.18	&........      & 298    & 37\\
HD26965     &0.82       &37.10     &-5.59	&$>$9280      &  4750   & 750\\
HD30495     &0.64       & 7.60     &-4.86	&   6080      &  428    & 58\\
HD32147     &1.06       &47.40     &-5.87	&$<$5450      &  4510   & 700\\
HD35296     &0.53       & 5.00     &-4.52	&........      & 388    & 66\\
HD36435     &0.78       &11.20     &-4.90	&........      & 531    & 66\\
HD38392     &0.94       &17.30     &-4.77	&........      & 816    & 100\\
HD39587     &0.59       & 5.20     &-4.51	&   4320      &  270    & 38\\
HD42807     &0.66       & 7.80     &-4.83	&........      & 410    & 54\\
HD43834     &0.72       &32.00     &-6.05	&   8760      &  4900   & 800\\
HD52698     &0.90       &26.00     &-4.74	&........      & 1960   & 280\\
HD53143     &0.81       &16.40     &-4.67	&........      & 1010   & 130\\
HD72905     &0.62       & 4.10     &-4.47	&........      & 144    & 18\\
HD75332     &0.52       & 4.00     &-4.35	&   1880      &  277    & 48\\
HD76151     &0.67       &15.00     &-5.24	&   1320      &  1380   & 200\\
HD78366     &0.60       & 9.70     &-4.75	&$<$680      &   850    & 130\\
HD81809     &0.64       &40.20     &-6.25	&........ & 10600\tablenotemark{A} & 1900\\
HD82106     &1.00       &13.30     &-4.64	&$<$600      &   435    & 50\\
HD95735     &1.51       &48.00     &-5.12	&........      & 2530   & 370\\
HD97334     &0.60       & 7.60     &-4.51	&$<$2920      &  529    & 77\\
HD98712     &1.36       &11.60     &-4.08	&........      & 194    & 20\\
HD101501    &0.74       &16.00     &-5.17	&$>$11320      & 1200   & 170\\
HD114613    &0.70       &33.00     &-5.85	&   5200      &  5600   & 930\\
HD114710    &0.58       &12.40     &-5.50	&$<$1120      &  1530   & 250\\
HD115383    &0.58       & 3.30     &-4.82	&$<$760      &   120    & 16\\
HD115404    &0.92       &19.00     &-5.25	&........      & 1020   & 130\\
HD128620    &0.71       &29.00     &-6.45	&   7840      &  4200   & 670\\
HD128621    &0.88       &42.00     &-5.97	&$>$11360      & 5170   & 820\\
HD131156A   &0.72       & 6.20     &-4.70	&$<$760      &   207    & 24\\
HD131977    &1.11       &44.60     &-5.38	&$<$600      &   3690   & 560\\
HD141004    &0.60       &18.00     &-6.18	&   6320      &  2780   & 460\\
HD147513    &0.62       & 8.50     &-4.61	&$<$680      &   587    & 84\\
HD147584    &0.55       &13.00     &-4.58	&........      & 2080   & 370\\
HD149661    &0.81       &23.00     &-4.96	&$<$4160      &  1950   & 280\\
HD152391    &0.75       &11.10     &-4.66	&   720      &   574    & 73\\
HD154417    &0.58       & 7.60     &-4.91	&  4200      &   597    & 91\\
HD155885    &0.86       &21.11     &-4.71	&........      & 1440   & 200\\
HD155886    &0.85       &20.69     &-4.65	&........      & 1420   & 190\\
HD156026    &1.16       &18.00     &-5.23	&$<$480      &   593    & 71\\
HD160346    &0.96       &36.00     &-5.36	&........      & 3210   & 480\\
HD165185    &0.62       & 5.90     &-4.43	&........      & 291    & 39\\
HD165341    &0.86       &19.70     &-5.18	&........      & 1260   & 170\\
HD166620    &0.87       &42.00     &-6.19	&$>$11200      & 5300   & 845\\
HD176051    &0.59       &16.00     &-5.70	&........      & 2350   & 390\\
HD185144    &0.79       &29.00     &-5.58	&........      & 3220   & 490\\
HD187691    &0.55       &10.00     &-5.97	&   3200      &  1250   & 210\\
HD190007    &1.12       &29.30     &-5.01	&$<$1760      &  1620   & 220\\
HD190406    &0.61       &14.50     &-5.58	&   3160      &  1730   & 270\\
HD191408    &0.87       &45.00     &-6.42	&$>$7640      &  6050   & 980\\
HD194012    &0.51       & 7.00     &-5.49	&........      & 900    & 170\\
HD201091    &1.17       &37.90     &-5.51	&$<$440      &   2450   & 350\\
HD201092    &1.37       &48.00     &-5.32	&$<$680      &   2960   & 440\\
HD206860    &0.58       & 4.70     &-4.62	&$<$880      &   237    & 34\\
HD209100    &1.06       &22.00     &-5.69	&........      & 1030   & 130\\
HD216803    &1.10       &10.30     &-4.54	&$<$520      &   223    & 23\\
HD219834B   &0.91       &42.00     &-5.49	&$>$13200      & 4820   & 760\\
HD224930    &0.67       &33.00     &-5.90	&........     & 6330   & 1100\\

\enddata


\tablenotetext{A}{This gyro age should be treated with caution because this star is a spectroscopic binary (see text, and Pourbaix, 2000).}


\end{deluxetable}

\section{Ages via gyrochronology for components of wide binaries}

As we have seen, testing these (or other) stellar ages is complicated because 
no star apart from the Sun has an accurately determined age. However, it is
possible to test the ages in a relative manner by asking whether the individual 
components of binary stars yield the same age. This test has been applied, with
mixed results, to chromospheric ages by Soderblom et al. (1991) and 
Donahue (1998).
We show here that gyrochronology yields substantially similar ages for both 
components of the three main sequence wide binary systems where {\it measured} 
rotation periods are available for the individual stars. 
[This latter requirement excludes otherwise interesting systems like 
16\,Cyg\,A/B (HD186408/HD186427; e.g. Cochran et al. 1997), where the rotation 
periods are derived quantities (Hale 1994), and 70\,Oph\,A/B (HD165341A/B), in 
the Mt.\,Wilson sample, where only the A component has a measured period 
(Noyes et al. 1984; Baliunas et al. 1996).]


\subsection{$\xi$\,Boo\,A/B (HD131156A/B)}
$\xi$\,Boo\,A/B is a wide main sequence binary (G8V+K4V) in the Mt.\,Wilson 
sample. The orbit calculated by Hershey (1977) gives a period of 152\,yr and 
eccentricity of 0.51, suggesting no rotational interaction between the 
components.
The Mt.\,Wilson datasets (Noyes et al. 1984; Baliunas et al. 1996; Donahue et
al. 1996) 
provide separate color and period measurements for both components, 
making the system particularly valuable as a test of the mass dependence of
rotation, under the assumption that binarity does not affect their rotation. 
Since the components of binaries are usually considered to be coeval, 
gyrochronology ought to give the same age for the individual components. 
For this system, gyrochronology yields ages of 187\,Myr and 265\,Myr for the 
bluer and redder components respectively (Table\,7), which gives a formal mean 
age for the system of 226\,$\pm$18Myr. 
The individual values, though not in agreement within the formal errors, 
are closer together than those provided by other methods.
For example, the chromospheric ages for the components are 232\,Myr and 
508\,Myr respectively, which also suggest a young age for the system.
As regards isochrone ages, Fernandes et al. (1998) have derived 
an isochrone age for the system of $2 \pm 2$Gyr. 
More recently, Takeda et al. (2007) have derived isochrone ages
for the A and B components of $<$0.76\,Gyr and $>$12.60\,Gyr respectively, 
attesting to the difficulty of applying the isochrone method to field stars.

\subsection{61\,Cyg\,A/B (HD201091/HD201092)}
There is a second, lower mass, main sequence wide binary (K5V+K7V) in the 
Mt.\,Wilson sample for which measured colors and periods are available. 
This is the 61\,Cyg\,A/B visual binary system, whose parameters (from Donahue
et al. 1996, see also Baliunas et al. 1996 and Hale, 1994) 
are listed in Table\,7. 
The orbit from Allen et al. (2000) suggests a semi-major axis of 85.6 AU and 
eccentricity of 0.32, while that from Gorshanov et al. (2005) suggests a period 
of 659yr and eccentricity of 0.48. 
Neither of these suggest an interaction between the components.
Gyrochronology yields ages of 2.12\,Gyr and 1.87\,Gyr for the A and B 
components respectively, suggesting a mean age for the system of 
2.0$\pm$0.2\,Gyr (see Fig.\,16), where the large differential rotation of the 
components contributes significantly to the error.
The corresponding chromospheric ages for the same stars are 2.36\,Gyr and 
3.75\,Gyr respectively, again in reasonable agreement, but not as close as the
gyro ages. 
The isochrone ages for these stars, upper limits of $<$0.44\,Gyr and 
$<$0.68\,Gyr respectively (Takeda et al. 2007), seem somewhat short. 

\subsection{$\alpha$CenA/B (HD128620/HD128621)}
We now consider the famous older system $\alpha$CenA/B, its G2V and K1V
components bracketing the Sun in mass, a system much studied by many
researchers over the years (e.g. Guenther and Demarque, 2000; 
Miglio \& Montalban 2005), and of special interest to asteroseismologists. 
Heintz (1982) has calculated an orbit with period of $\sim$80yr and eccentricity
of 0.516, suggesting that the components have not suffered rotational 
interactions.
The published ages for the system range from 4\,Gyr to 8\,Gyr, depending on the 
details of the models used (see, e.g. Guenther \& Demarque 2000).
Guenther and Demarque (2000) themselves derive an age range of 7.6-6.8\,Gyr, 
somewhat older than the Sun, depending on whether or not $\alpha$CenA has a 
convective core. Eggenberger et al. (2004) suggest an age of 6.5$\pm$0.3\,Gyr.
Using the rotation periods provided by Ed Guinan (2006, personal communication),
28$\pm$3d and 36.9$\pm$1.8d for the A and B components respectively
\footnote{Pizzolato et al. (2003) lists periods of 29d and 42d respectively, sourced from Saar and Osten (1997), which in turn sources the first to Hallam, Aliner \& Endal (1991), and states that the latter is estimated from CaII measurements.}, 
and $B-V$ colors\footnote{SAB thanks David Frew for his trouble researching these colors.} 
of 0.67$\pm$0.02 and 0.87$\pm$0.02, we 
derive ages for the components of 4.6\,Gyr and 4.1\,Gyr, with a mean of
4.4$\pm$0.5\,Gyr, toward the lower end of the published ages\footnote{The Pizzolato et al. (2003) periods would yield a slightly older gyro age of 4.6\,Gyr for the system.}, 
but in good agreement with one another\footnote{There is a third component in the $\alpha$Cen system, $\alpha$Cen\,C (Proxima Centauri), and it too has a measured period, 31$\pm$2d, but its spectral type is M5V, so it is not on the interface sequence (and hence not considered here), and it ought to follow the age dependence appropriate for the C sequence stars, but this dependence is not yet known well.}. 
These stars, and the corresponding isochrone are also plotted in Fig.\,16.
The chromospheric ages for the $\alpha$Cen\,A and B components using $R'_{HK}$ 
values from Henry et al. (1996) are 5.62 and 4.24\,Gyr respectively, again 
comparable, if not as close. 
In comparison, the isochrone ages from Takeda et al. (2007) for the A and B 
components, derived separately, are 7.84\,Gyr and $>$11.36\,Gyr respectively.

\subsection{36\,Oph\,A/B/C (HD155886/HD155885/HD156026)}
Finally, we consider the triple system 36\,Oph\,A/B/C, included in the 
Mt.\,Wilson sample. A and B are two chromospherically active K1 dwarfs, while 
the distant tertiary, C, is a K5 dwarf. The AB orbit has a period of 
$\sim$500yr, but a very high eccentricity of $\sim$0.9, implying a closest 
approach of A and B of order 6AU (Brosche 1960; Irwin et al. 1996).
The latter fact suggests proceeding with caution, because A and B could
potentially have interacted rotationally. 

We have used the observed periods of 20.69d, 21.11d, and 18.0d, listed in 
Donahue et al. (1996) and Baliunas et al. (1983) for the A, B, and C components 
respectively to plot these in the color-period diagram displayed in Fig.\,17\footnote{Baliunas et al (1996) list a joint period of 21d for all three components. Pizzolato et al. (2003) reference Saar \& Osten (1997) for the 20.69d and 21.11d periods for A and B, and Hempelmann et al. (1995), who in turn references Noyes et al. (1984) for the 18.0d (observed) period for C. Saar \& Osten (1997) themselves reference Donahue et al. (1996) for the A and B periods, and say that the 18.5d period is estimated from CaII measurements.}.
The gyro ages for A and B are both nominally 1.43\,Gyr, but that for C is only 
590$\pm$70\,Myr. 
We favor the lower age here because the C component is distant, while 
the A and B components seem to have interacted and presumably spun down to 
their $\sim$21d periods from the $\sim$13.4d periods that would otherwise be 
expected for the 590\,Myr age for the system.

Interestingly, the chromospheric ages for the A, B, and C components range from 
1.1\,Gyr to 1.4\,Gyr, similar to the gyro age for the A/B pair.
The isochrone age for the C component only, provided by 
Takeda et al. (2007) is $<$480\,Myr, again suggesting a youthful system. 
The fact that the A and B components have essentially the same mass provides a
simplification that could be quite useful to further studies of this system.

For the present state of gyrochronology, we consider the particular cases
presented above to represent success.


\clearpage

\begin{deluxetable}{rlrrrr}
\tabletypesize{\scriptsize}
\tablecaption{Ages for wide binary systems. \label{tbl-7}}
\tablewidth{0pt}
\tablehead{
\colhead{Star} & \colhead{$B-V$}   & \colhead{$\bar{P}_{rot}$(d)\tablenotemark{a}} &\colhead{$Age_{chromo}$} & \colhead{$Age_{iso}$} & \colhead{$Age_{gyro}$\tablenotemark{b}}   
}
\startdata
HD131156A & 0.76 & 6.31(0.05)  & 232\,Myr &$<$760\,Myr    & 187$\pm$21\,Myr\\
HD131156B & 1.17 & 11.94(0.22) & 508\,Myr &$>$12600\,Myr  & 265$\pm$28\,Myr\\
Mean       &      &           &          &         & {\bf 226$\pm$18\,Myr}\\
\hline
HD201091 &1.18  &35.37(1.3) & 2.36\,Gyr &$<$0.44\,Gyr  & 2.12$\pm$0.3\,Gyr\\
HD201092 &1.37  &37.84(1.1) & 3.75\,Gyr &$<$0.68\,Gyr  & 1.87$\pm$0.3\,Gyr\\
Mean      &       &            &           &     & {\bf 2.0$\pm$0.2\,Gyr}\\
\hline
HD128620 & 0.67 & 28(3)   & 5.62\,Gyr   &7.84\,Gyr     & 4.6$\pm$0.8\,Gyr\\
HD128621 & 0.87 & 36.9(1.8) & 4.24\,Gyr &$>$11.36\,Gyr & 4.1$\pm$0.7\,Gyr\\
Mean     &      &            &            &     & {\bf 4.4$\pm$0.5\,Gyr}\\
\hline
\hline
HD155886 & 0.85 & 20.69(0.4) & 1.1\,Gyr  & ......      & 1.42$\pm$0.19\,Gyr\\
HD155885 & 0.86 & 21.11(0.4) & 1.2\,Gyr  & ......      & 1.44$\pm$0.20\,Gyr\\
HD156026 & 1.16 & 18.0(1.0) & 1.4\,Gyr &$<$0.48\,Gyr &{\bf 0.59$\pm$0.07\,Gyr}\\
\enddata


\tablenotetext{a}{Differential rotation is the main contributor to the period errors in the parentheses.}
\tablenotetext{b}{Boldface figures denote the final gyro age for each system}


\end{deluxetable}

\section{Comparison with isochrone ages}

A uniform comparison of gyro- and isochrone ages was not possible until 
Takeda et al. (2007) submitted a manuscript to the Astrophysical Journal
Supplement subsequent to this submission. 
This paper contains a very careful derivation of isochrone ages for the 
$\sim$1000 stars in the Spectroscopic Properties of Cool Stars Catalog (SPOCS).
This catalog (Valenti \& Fischer 2005) itself consists of high-resolution
echelle spectra and their detailed analysis of over 1000 nearby F-, G- and 
K-type stars obtained through the Keck, Lick, and Anglo-Australian Telescope 
planet search programs, including $\sim$100 stars with known planetary 
companions.

Takeda et al. (2007) conduct a Bayesian analysis of the stellar parameters 
using reasonable priors to generate a probability distribution function (PDF) 
for the age of each star. This method permits the identification of a 
`well-defined' age for a star if the PDF peaks appropriately, or just as
importantly, the derivation of an isochrone upper- or lower limit for the age.
Indeed, for most of the stars common to our sample and theirs, they derive
only such a limit, as a glance at the column for isochrone ages in Tables 3, 5 
and 6 shows. However, for 26 of these (common) stars, Takeda et al. (2007) list 
well-defined ages, and these can be compared to the corresponding gyro ages.

This comparison is shown in Fig.\,18. Of these 26 stars, only 3 lie above the 
line of equality, and 13 have isochrone ages within a factor of two of 
the gyro ages, all higher than the corresponding gyro ages. In fact, the median 
isochrone age is 2.7 times the median gyro age. Evidently, the Bayesian 
technique used still does not eliminate the known 
bias in the isochrone ages towards older values. 

In fact, the same test applied to the binary systems in the previous
section with respect to gyro ages yields uncertain results with respect to 
these isochrone ages. Indeed, of the 9 stars under consideration, only one 
($\alpha$\,Cen\,A) has a well-defined isochrone age, and the rest upper- or 
lower limits. These stars are also plotted in Fig.\,18, with dashed lines 
joining the binary components, and arrows indicating upper- or lower limits.   

In summary, it would seem that the isochrone ages are still problematical,
despite the careful analysis of Takeda et al. (2007). Of course, as we have 
noted in the introduction, it is perhaps not fair and evidently not possible, 
to use slowly varying parameters to derive precise ages for stars on the main 
sequence. The two methods are, however, complementary in that it might be 
preferable to use gyro ages on the main sequence, and isochrone ages off it.

\section{Conclusions}

The rotation period distributions of solar and late-type stars suggest that 
coeval stars are preferentially located on one of two sequences.
The mass- and age dependencies of one of these sequences, the interface 
sequence, are shown to be universal, shared by both cluster and field stars, 
and we have specified them using simple functions, generalizing the dependence
originally suggested by Skumanich (1972). The mass dependence is derived 
observationally using a series of open clusters, and the age dependence,
roughly $\sqrt t$, is specified via a solar calibration.


The dependencies are inverted to provide the age of a star as a function 
of its rotation period and color, a procedure we call gyrochronology. 
Errors are calculated for such ages, based on the data currently available,
and shown to be roughly 15\% (plus possible systematic errors)  for 
individual stars.
Because the dependencies are universal, they must also apply to field stars,
but the derivation of such ages requires excising pre-I\,sequence stars, 
facilitated by their location below the I\,sequence in color-period diagrams. 
The short lifetime of this pre-I\,sequence phase assures us that all such stars 
are less than a couple of hundred million years in age.

Using this formalism, we have calculated ages via gyrochronology for individual
stars in three illustrative groups of field stars, and listed them along with
the errors. 
For the first group, the Mt.\,Wilson stars, these ages are shown to be in 
general agreement with the chromospheric ages, except that stars bluer than the 
Sun have systematically higher chromospheric ages, the median chromospheric age 
being higher by about 33\%.
The majority of the second group, from Strassmeier et al. (2000), are shown to 
be younger than 1\,Gyr, in keeping with the selection of the sample
for activity, which correlates negatively, as expected, with gyro age. 
The third group, from Pizzolato et al. (2003), are shown to be somewhat older, 
partially due to an overlap with the Mt.\,Wilson sample, and their X-ray fluxes
are shown to decay systematically with gyro age.
We have shown that gyrochronology yields similar ages for both components of 
three wide binary systems, $\xi$\,Boo\,A/B, 61\,Cyg\,A/B, and 
$\alpha$\,Cen\,A/B.
The 36\,Oph\,A/B/C triple system shows signs of rotational interaction between
the A and B components.
Finally, the recent Takeda et al. (2007) isochrone ages appear to be inferior 
to the gyro ages for the same main sequence stars.

Thus, we have re-investigated the use of a rotating star as a clock, 
clarified and improved its usage, calibrated it using the Sun,
and demonstrated that it keeps time well.



\acknowledgments

The word ``gyrochronology'' was inspired by the work of A.\,E.\,Douglass 
on dendrochronology at Lowell Observatory.
SAB would like to acknowledge Sabatino Sofia as a constant source of 
intellectual and moral support and many discussions, and Charles Bailyn for 
initially suggesting the removal of the age dependence. Marc Buie, Will 
Grundy, Wes Lockwood, Bob Millis, Byron Smith, Brian Skiff and my other 
colleagues at Lowell have supported me in numerous ways. Stephen Levine read 
the manuscript closely, and found an algebraic error.
David James, Heather Morrison, Steve Saar, Sukyoung Yi and an anomymous referee 
are gratefully acknowledged for input on a prior version of the paper.
The paper owes much to the baristas at Late For The Train, Flagstaff.
Finally, this material is based upon work partially supported by the National 
Science Foundation under Grant No. AST-0520925.

\appendix
\section{Appendix: Derivation of the error on the index $n$}

By definition,
\begin{equation}
{
P = f(B-V).g(t) = a\,x^b\,t^n
}
\end{equation}
Taking natural logarithms and rearranging, we get 
\begin{equation}
{
n = \frac{ln\,P_{\odot} - ln\,a - b\,ln\,x_{\odot}}{ln\,t_{\odot}} = \frac{U}{V}
}
\end{equation}
Differentiating yields
\begin{equation}
{
\frac{dn}{n} =  \frac{dU}{U} - \frac{dV}{V} 
}
\end{equation}
or
\begin{equation}
{
\frac{dn}{n} =    \frac{1}{U} [\frac{dP_{\odot}}{P_{\odot}} - \frac{da}{a} 
              - b \frac{dx_{\odot}}{x_{\odot}} - ln\,x_{\odot}\,db]  
              -   \frac{dt_{\odot}}{t_{\odot}\,ln\,t_{\odot}}
}
\end{equation}
Adding the errors in quadrature yields
\begin{equation}
{
(\frac{\delta n}{n})^2 =  (\frac{\delta t_{\odot}}{t_{\odot}\,ln\,t_{\odot}})^2
                        + \frac{1}{U^2} [  (\frac{\delta P_{\odot}}{P_{\odot}})^2 
                                         + (\frac{\delta a}{a})^2
                                         + (b\frac{\delta x_{\odot}}{x_{\odot}})^2
                                         + (ln\,x_{\odot}\,\delta b)^2
                                         ]
}
\end{equation}
For the error in the age of the Sun (4566\,Myr; $ln\,t_{\odot}=8.426$), 
we adopt the value of 50\,Myr\footnote{Allegre et al. (1995) list the 
impressively small error of $+2/-1$\,Myr 
(in agreement with the present day precision of radioactive dating techniques)
for the age of the formation of the Allende refractory inclusions, 
generally accepted as the age of the Earth/meteorites/Solar System. However,
we astronomers do not know what event in the Sun's history corresponds to this 
point. Is this the zero age main sequence, or the birthline 43\,Myr earlier 
(Barnes \& Sofia, 1996), or some other event entirely? In view of these 
uncertainties, we adopt an error of 50\,Myr in the age of the Sun.},
for that in the rotation period, 1\,d (consistent with the measured range 
in the solar rotation period - see section 4 and Donahue et al. 1996),
and for that in the solar $B-V$ color ($x = B-V_{\odot} = 0.242$), 
we adopt the value 0.01.
From section\,2, $a = 0.7725 \pm 0.011$ and $b = 0.601 \pm 0.024$.
Input of these values yields
\begin{equation}
{
(\frac{\delta n}{n})^2 =  (\frac{50}{4566 \times 8.43})^2
                        + \frac{1}{4.37^2} [  (\frac{1}{26.09})^2 
                                         + (\frac{0.011}{0.7725})^2
                                         + (0.601\frac{0.01}{0.242})^2
                                         + (-1.419 \times 0.024)^2
                                           ]
}
\end{equation}
or
\begin{equation}
{
(\frac{\delta n}{n})^2 =  1.69 \times 10^{-6}
                                         + 10^{-6} [ 77.4 + 10.6 + 32.3 + 60.7 ]
                       = 182.6 \times 10^{-6}
}
\end{equation}
or\footnote{Note that the largest terms come from the differential rotation of
the sun and the index $b$, while the age error of the sun contributes little to 
the error in $n$.},
\begin{equation}
{
\frac{\delta n}{n} = 1.37 \times 10^{-2}
}
\end{equation}
so that
\begin{equation}
{
n =  0.5189 \pm 0.0070
}
\end{equation}
which shows that the index $n$ is determined well.






\clearpage 

\begin{figure}	   
\includegraphics[scale=1.25]{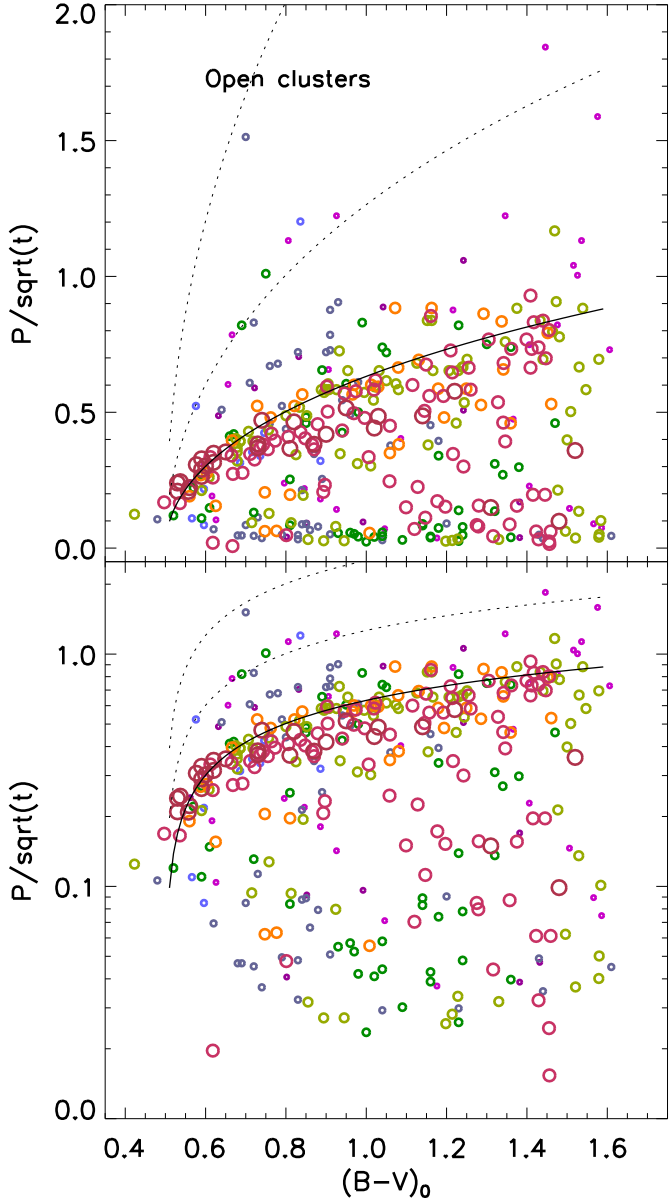}
\caption{Plot of $P/\sqrt{t}$ vs $(B-V)_0$ for open cluster stars only
($P$=rotation period; $t$=cluster age). 
Symbol sizes and colors correspond to cluster age (blue/small = young, 
red/large = old). The densest concentration of stars in the vicinity of the 
solid line represents the interface sequence. Note how the interface sequences 
of all the open clusters coincide. Also note the clearly visible convective 
sequence along the lower edge of the upper panel. The solid line represents 
$f(B-V)$. Dotted lines are at $2f$ and $4f$. Some stars in the vicinity of the 
dashed lines could be spurious periods or non-members.
The same data are plotted in both panels, on a linear scale in the upper panel, 
and on a logarithmic scale in the lower panel. 
\label{fig1}}
\end{figure}

\begin{figure}	
\includegraphics[scale=1.25]{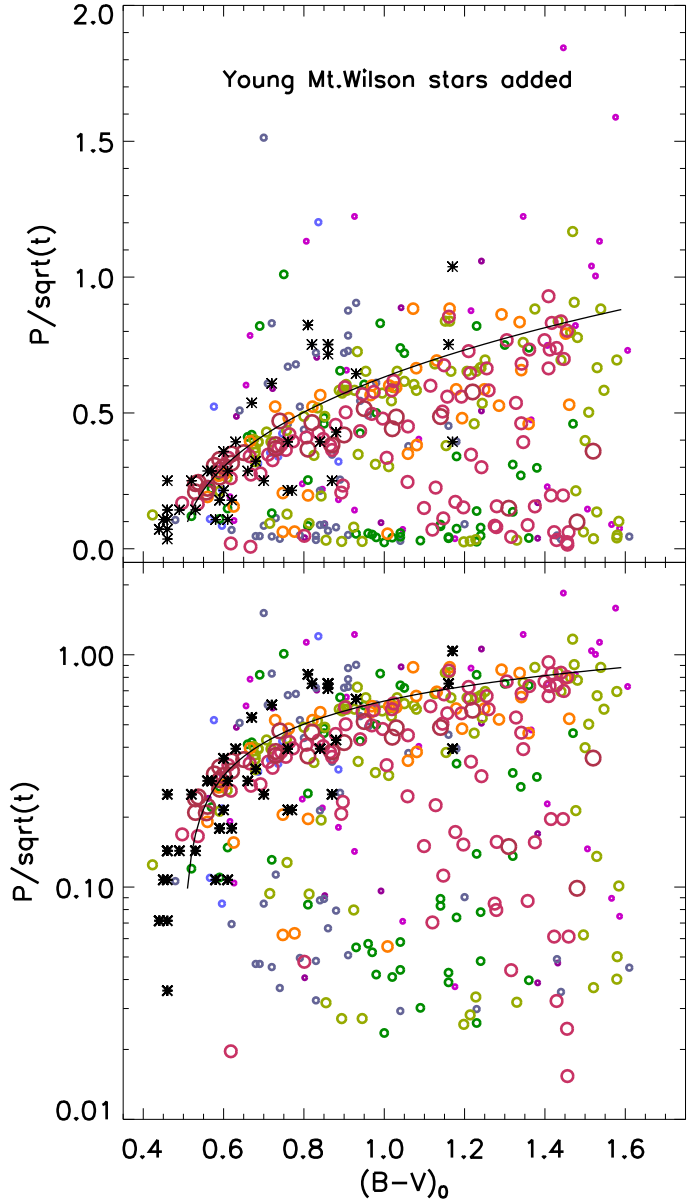}
\caption{Plot of $P/\sqrt{t}$ vs $(B-V)_0$ for the young Mt.\,Wilson stars 
(small black asterisks), assumed to be 780\,Myr old, the
median chromospheric age for this sample, overplotted on the open cluster data.
Note how the young Mt.\,Wilson stars overlie the interface sequences for the 
open clusters, and that no young Mt.\,Wilson stars are on the C sequence. 
The non-coeval nature of the young 
Mt.\,Wilson sample probably accounts for much of the dispersion observed.
\label{fig2}}
\end{figure}

\begin{figure}	
\includegraphics[scale=1.25]{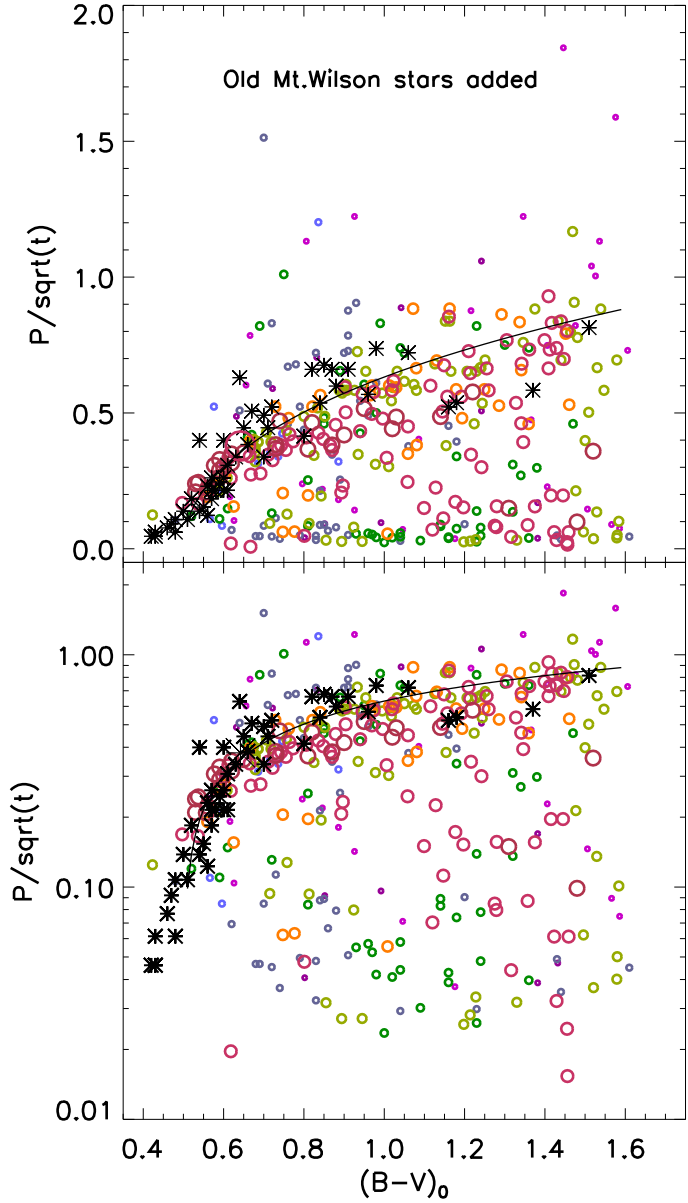}
\caption{Plot of $P/\sqrt{t}$ vs $(B-V)_0$ for the old Mt.\,Wilson stars 
(large black asterisks), assumed to be 4.24\,Gyr old, the 
median chromospheric age for this sample, overplotted on the open cluster data.
Note how the old Mt.\,Wilson stars overlie the interface sequences for the open 
clusters, and that no Mt.\,Wilson stars are located near the C sequence. 
The non-coeval nature of the old 
Mt.\,Wilson sample probably accounts for much of the dispersion observed.
\label{fig3}}
\end{figure}

\begin{figure}	
\includegraphics[scale=1.25]{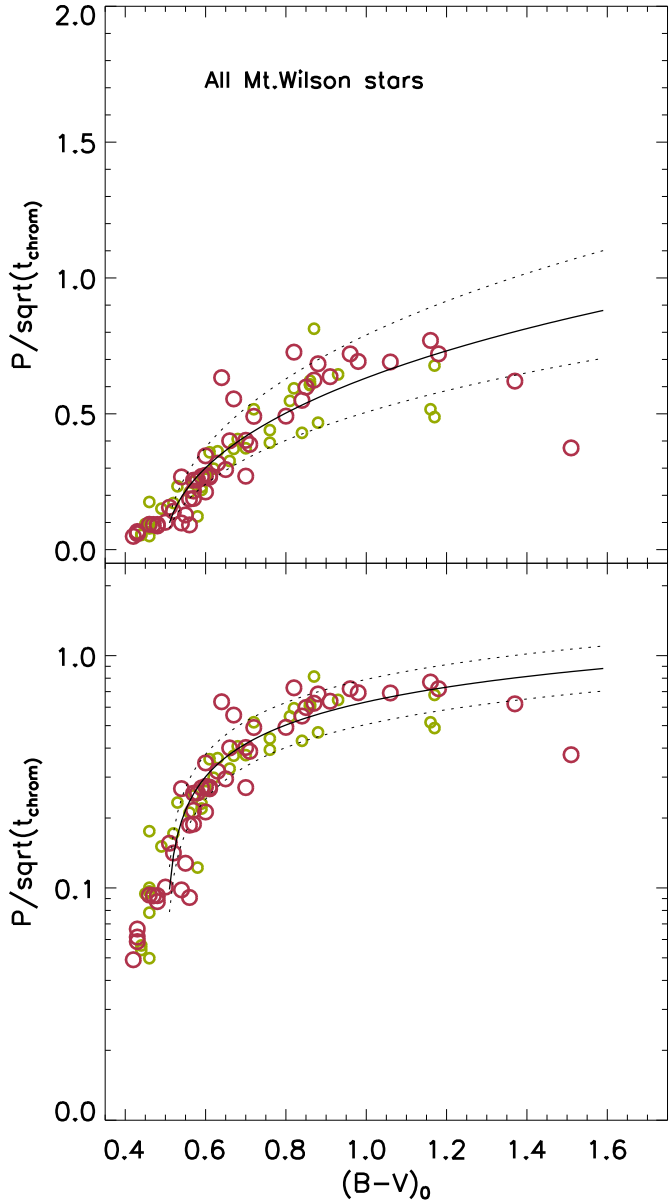}
\caption{Plot of $P/\sqrt{t}$ vs $(B-V)_0$ for the
individually age-corrected (chromospheric ages) Mt.\,Wilson stars.
Note how the Mt.\,Wilson stars (small green circles = young; 
large red circles =old) lie on top of the interface sequences 
for the open clusters. The solid line represents $f(B-V)$, as before, and the 
dotted lines are a factor of $0.8-$ and $1.25 \times f$ ($\pm$25\%).
\label{fig4}}
\end{figure}

\begin{figure}   
\plotone{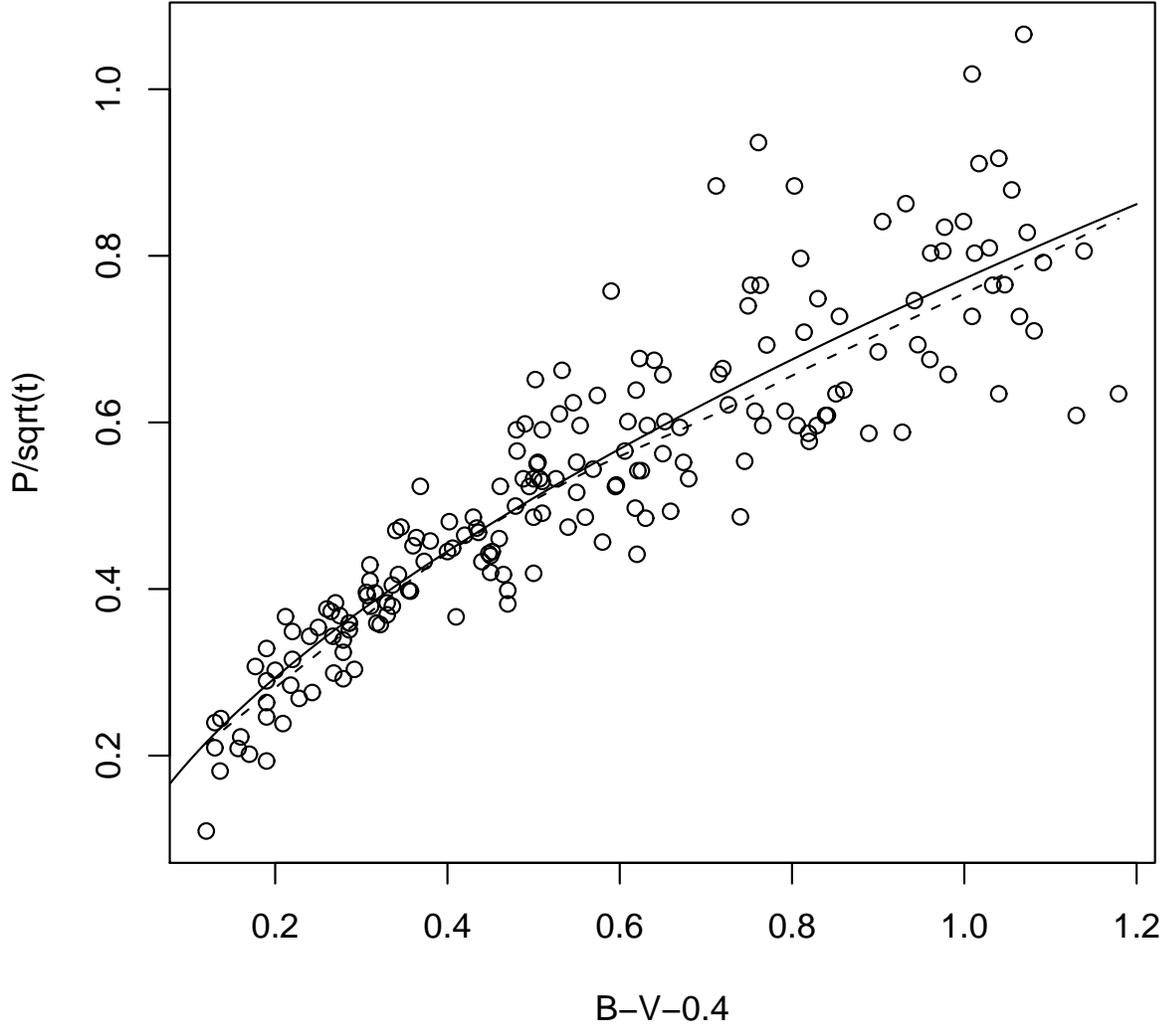}
\caption{The fit to the mass dependence (solid line), using R: 
$f(B-V)=(0.7725 \pm 0.011) \times (B-V_0-0.4)^{0.601 \pm 0.024}$.  
The abscissa gives $(B-V_o-0.4)$ and the ordinate $P/\sqrt{t}$ for individual 
I sequence stars in the main sequence open clusters listed in the text.
The dashed line shows a smooth trend curve plotted using the function 
{\it lowess} in the R statistics package. 
Note the similarity of the two curves, which demonstrates 
that the fitting function is appropriate for these data.
\label{fig5}}
\end{figure}

\begin{figure}	
\includegraphics[scale=1.25]{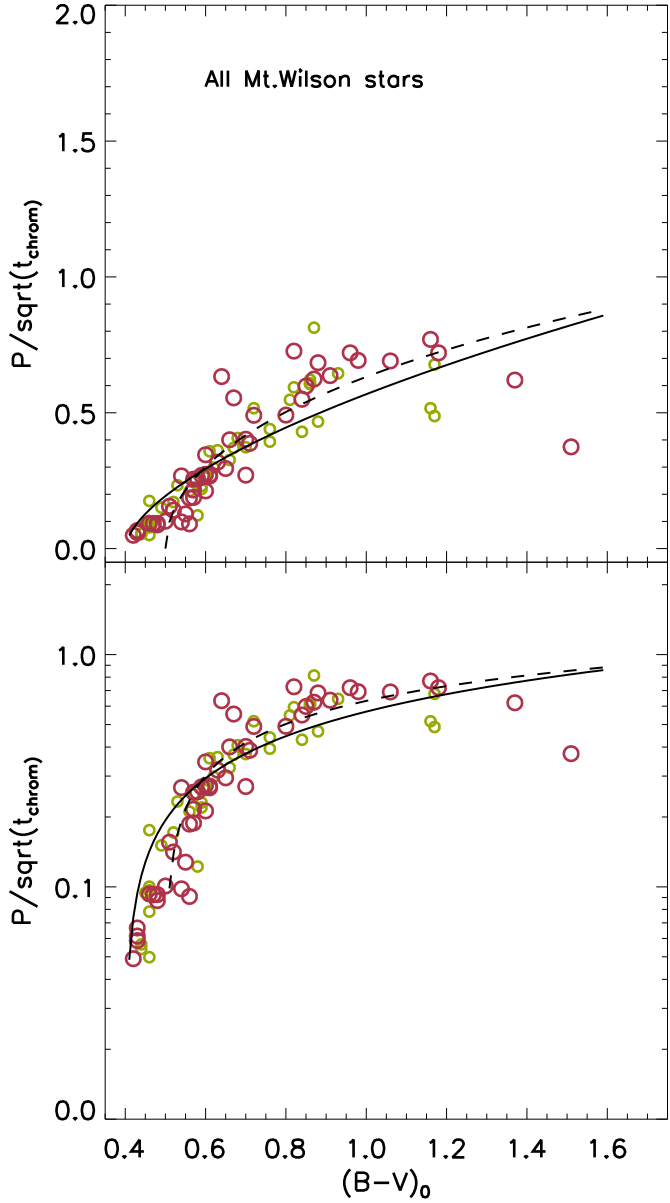}
\caption{Plot of $P/\sqrt(t)$ vs $(B-V)_0$ for individually age-corrected 
(chromospheric ages) Mt.\,Wilson stars (small green circles = young;
large red circles = old), with the old (dashed) and new (solid) functions, $f$, 
overplotted. Note that the new function accommodates bluer stars.
The discrepancy arises from the assumed chromospheric ages for the stars,
which are almost certainly overestimated for the F\,stars (see text).
The same data are plotted in both panels, on a linear scale in the 
upper panel, and on a logarithmic scale in the lower panel.
\label{fig6}}
\end{figure}

\begin{figure} 
\includegraphics[scale=1]{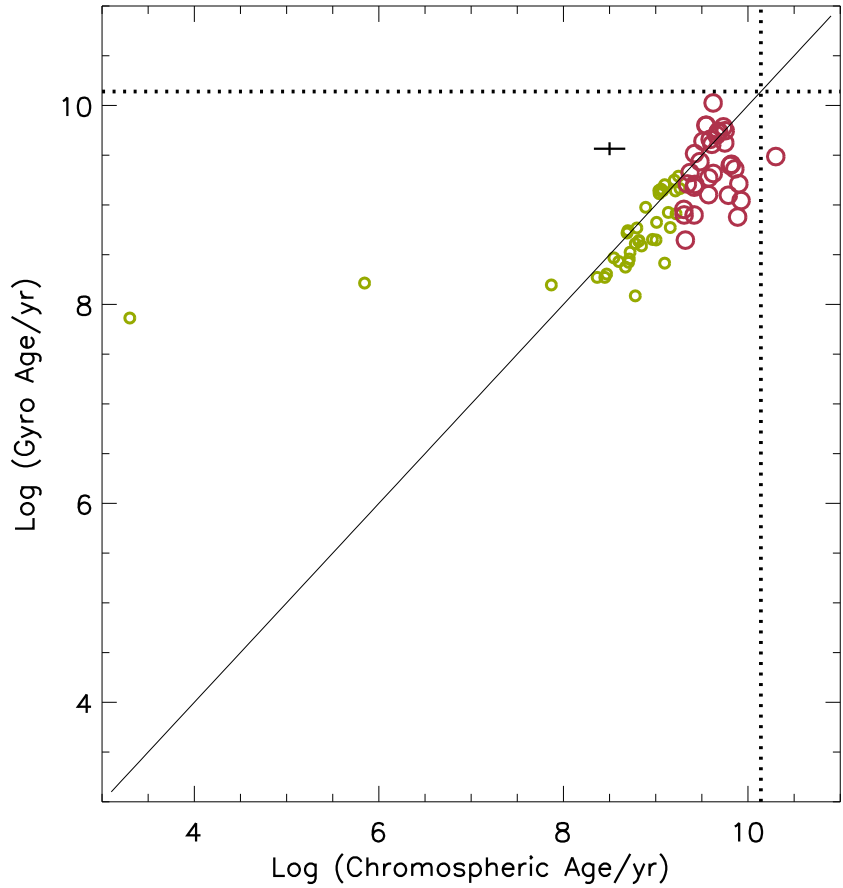}
\caption{Comparison of gyro- and chromospheric ages for the Mt.\,Wilson stars. 
The young (Y) and old (O) Mt.\,Wilson stars are marked with small green and 
large red circles respectively. The line indicates equality. Note that the 
gyro ages are well-behaved for the youngest stars, where the chromospheric 
ages are suspect. The dotted lines represent the age of the universe, and the
cross indicates typical gyro/chromospheric age errors quoted for this sample.
\label{fig7}}
\end{figure}

\begin{figure} 
\includegraphics[scale=1]{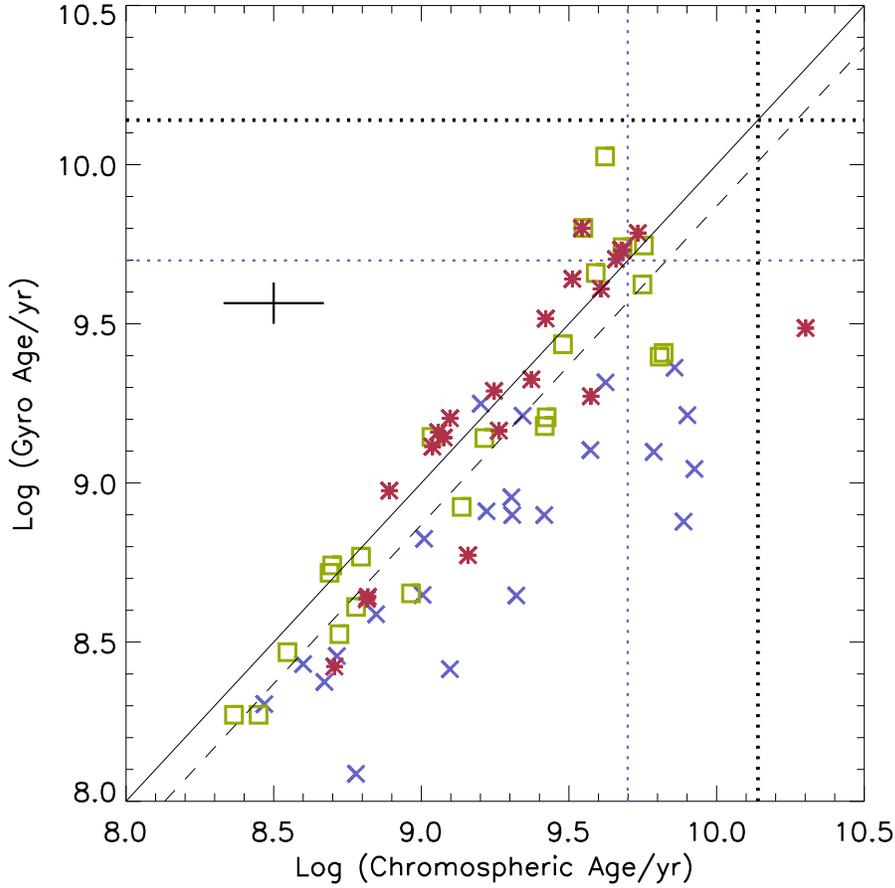}
\caption{Comparison of gyro and chromospheric ages for the Mt.\,Wilson stars. 
Blue crosses indicate stars bluer than $B-V=0.6$, red asterisks stars redder 
than $B-V=0.8$, and green squares those with colors between. 
The upper (solid) line indicates equality, while the lower (dashed) line at 
$Age_{gyro}=0.74 \times Age_{chromo}$ bisects the data.
Note that both techniques are in general agreement about the youth or antiquity
of any particular star, but that the gyro ages are roughly 25\% lower on 
average. The figure also shows that the bluer stars contribute most to this
discrepancy. The thick and thin dotted lines represent the age of the universe
and the lifetime of F\,stars (5\,Gyr) respectively.
The cross indicates typical gyro/chromospheric age errors quoted for this 
sample.
\label{fig8}}
\end{figure}

\begin{figure}   
\includegraphics[scale=1]{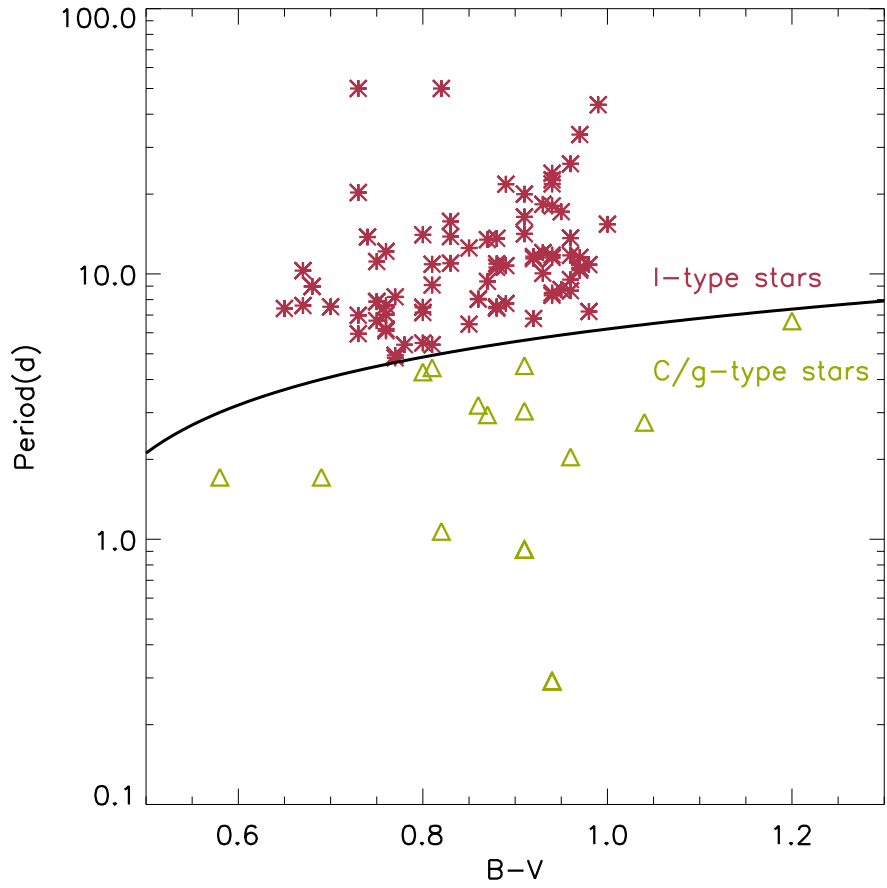}
\caption{ Division of the Strassmeier et al. (2000) sample into I sequence 
(suitable for gyrochronology) and C/g (unsuitable) categories.
The solid line separates the two categories of stars, and represents an
isochrone for 100\,Myr.
\label{fig9}}
\end{figure}

\begin{figure}   
\includegraphics[scale=1]{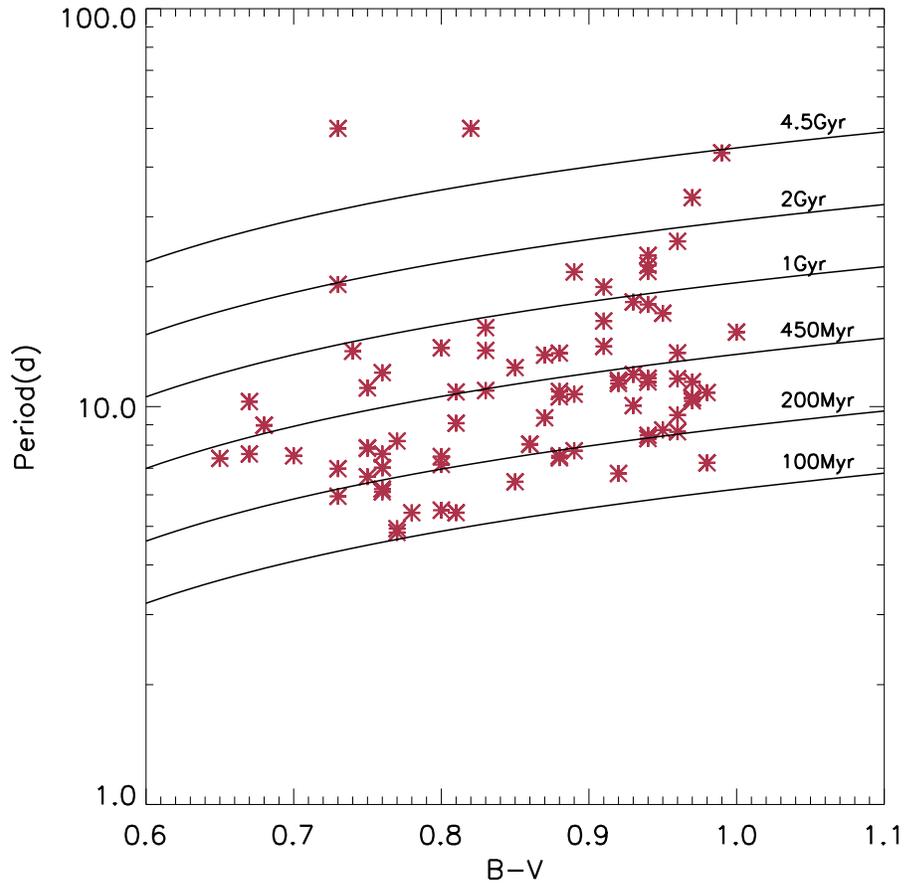}
\caption{ Ages for the Strassmeier et al. (2000) I sequence stars may be read 
off this figure. Isochrones correspond to ages of 100\,Myr, 200\,Myr, 450\,Myr,
1\,Gyr, 2\,Gyr, and 4.5\,Gyr. Note that all but 4 of the stars are less than
2\,Gyr in age. 
\label{fig10}}
\end{figure}

\begin{figure}   
\includegraphics[scale=1]{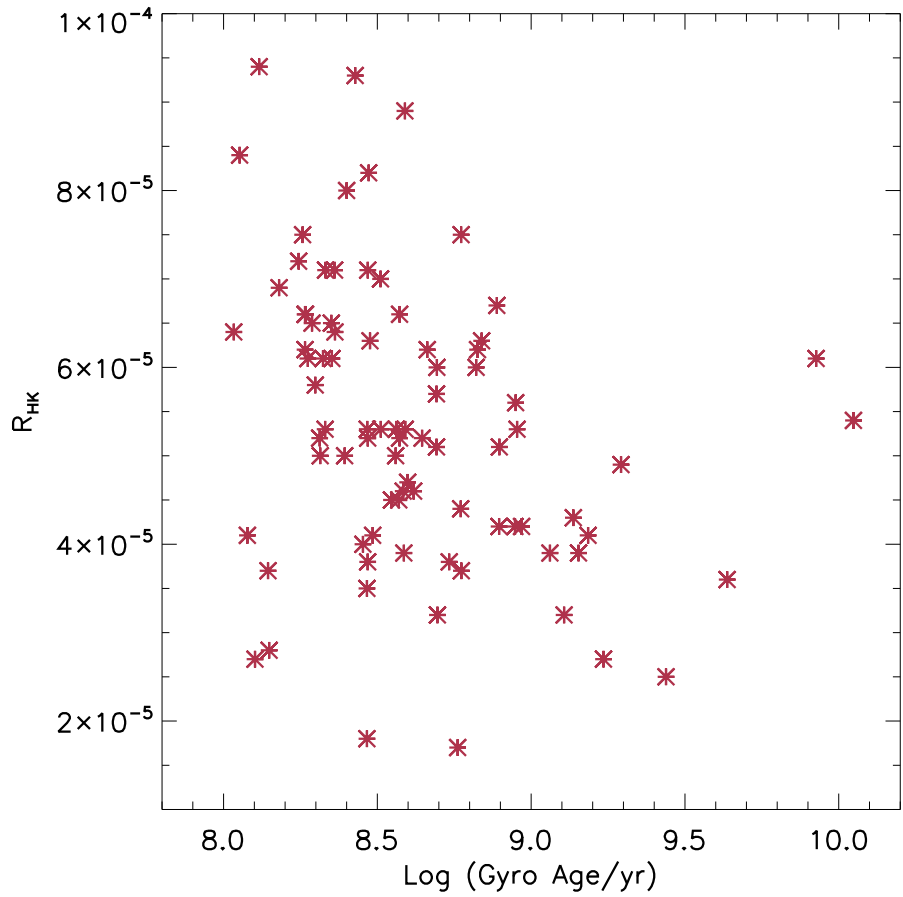}
\caption{ $R_{HK}$ vs. Gyro age for the Strassmeier et al. (2000) I sequence 
stars. Note the declining trend of $R_{HK}$ with age. The trend is obvious 
despite the fact that the $R_{HK}$ values are not long-term averages, and have 
not been corrected for photospheric contributions or variation with color. 
\label{fig11}}
\end{figure}

\begin{figure} 
\includegraphics[scale=1]{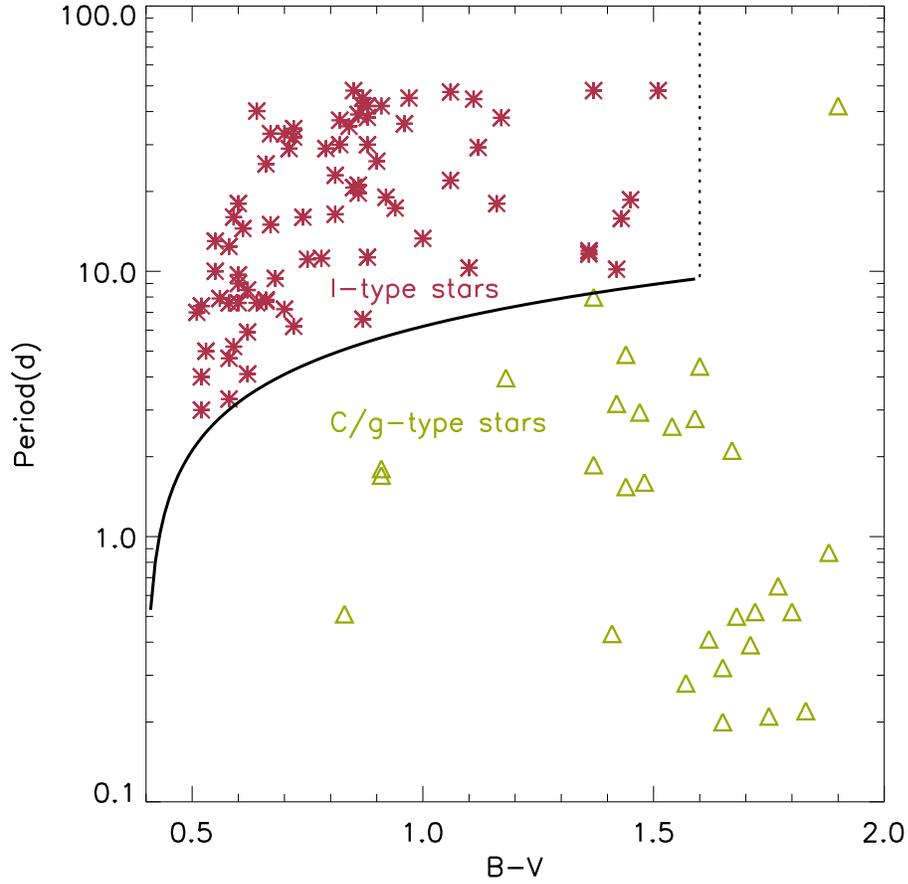}
\caption{ Division of the Pizzolato et al. (2003) stars into 
C/g- \& I\,categories.
The solid line separates the two categories of stars, and represents a 
rotational isochrone for 100\,Myr. The dotted line indicates the approximate
color ($B-V=1.6$; M\,3) for the onset of full convection.
\label{fig12}}
\end{figure}

\begin{figure} 
\includegraphics[scale=1]{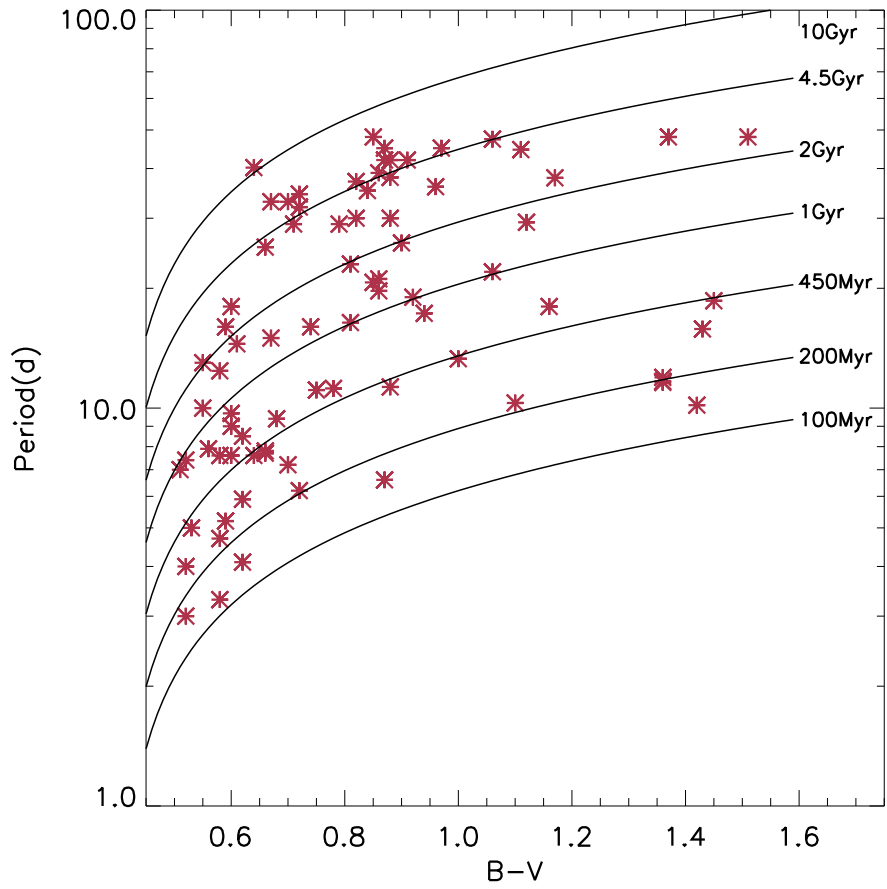}
\caption{ Ages for the Pizzolato et al. (2003) stars may be read off this
figure. Rotational isochrones correspond to ages of 
100\,Myr, 200\,Myr, 450\,Myr, 1\,Gyr, 2\,Gyr, 4.5\,Gyr, \& 10\,Gyr, as marked. 
\label{fig13}}
\end{figure}

\begin{figure} 
\includegraphics[scale=1]{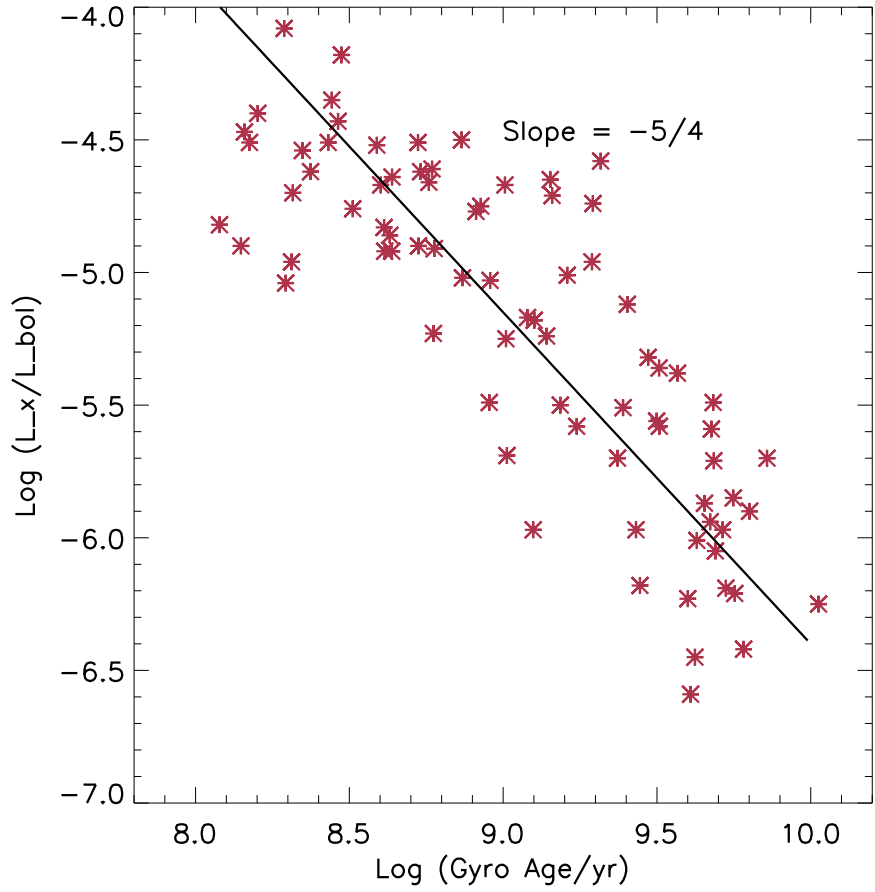}
\caption{ X-ray emission vs. gyro age for the Pizzolato et al. (2003) stars.
Note the steady decline in X-ray emission with gyro age, as expected.
The line drawn has a slope of $-5/4$, as expected from MHD turbulence.
\label{fig14}}
\end{figure}

\begin{figure} 
\includegraphics[scale=1.00]{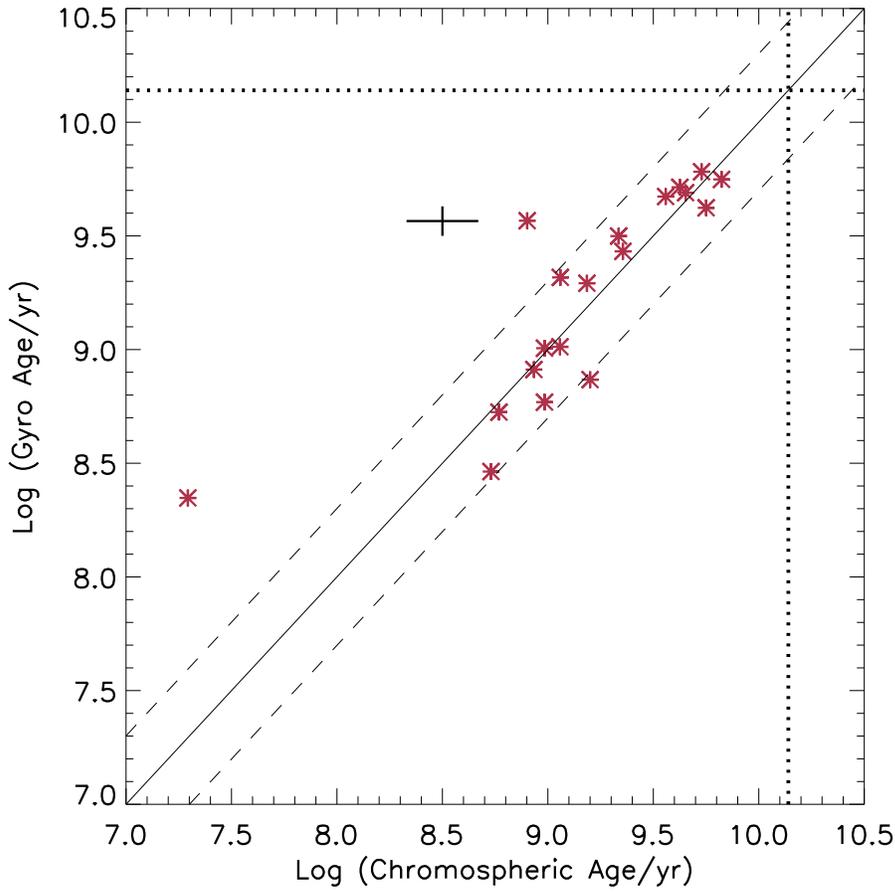}
\caption{Comparison of gyro and chromospheric ages for the 19 Pizzolato et 
al. (2003) stars in the Southern Chromospheric Survey of Henry et al. (1996). 
Note that almost all the stars scatter about the line of equality (solid).
The dashed lines indicate factors of two above and below the gyro ages. 
Typical error bars are indicated.
The dotted lines indicate the age of the universe.
\label{fig15}}
\end{figure}

\begin{figure}	
\includegraphics[scale=1.00]{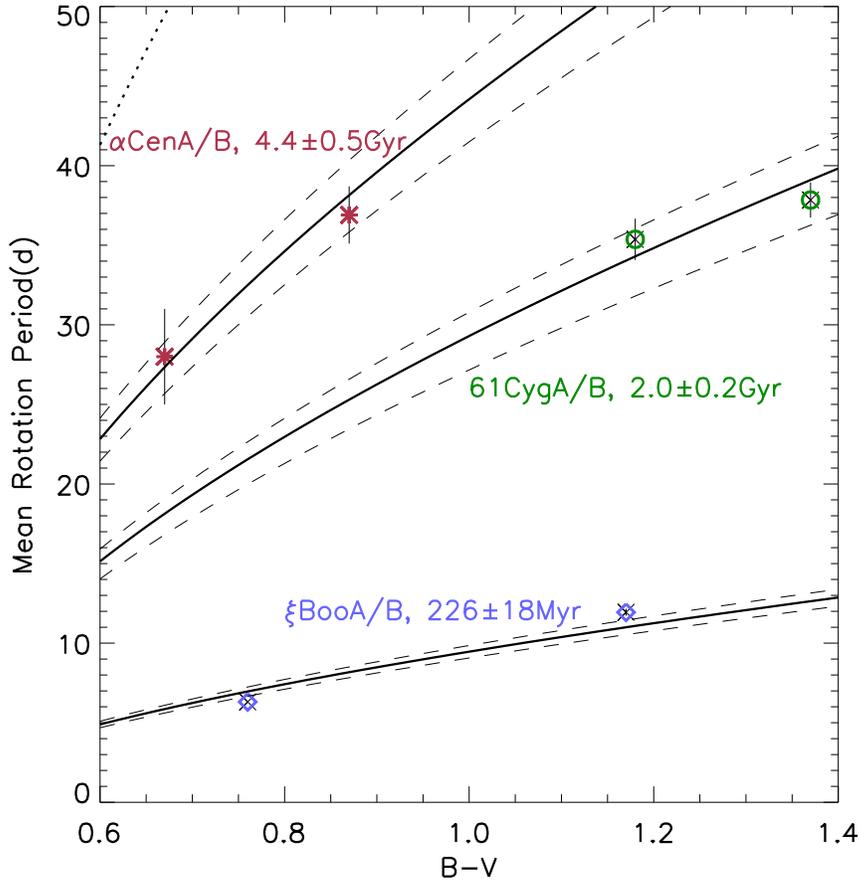}
\caption{Color-period diagram for three wide binary systems,
$\xi$\,Boo\,A/B, 61\,Cyg\,A/B, \& $\alpha$\,Cen\,A/B.
Rotational isochrones are drawn for ages of 226\,Myr, 2.0\,Gyr \& 4.4\,Gyr 
respectively, and the errors are indicated with dashed lines.
Note that for all three wide binary systems, both components give substantially 
the same age. 
The dotted line corresponds to the age of the universe.
\label{fig16}}
\end{figure}

\begin{figure}	
\includegraphics[scale=1.00]{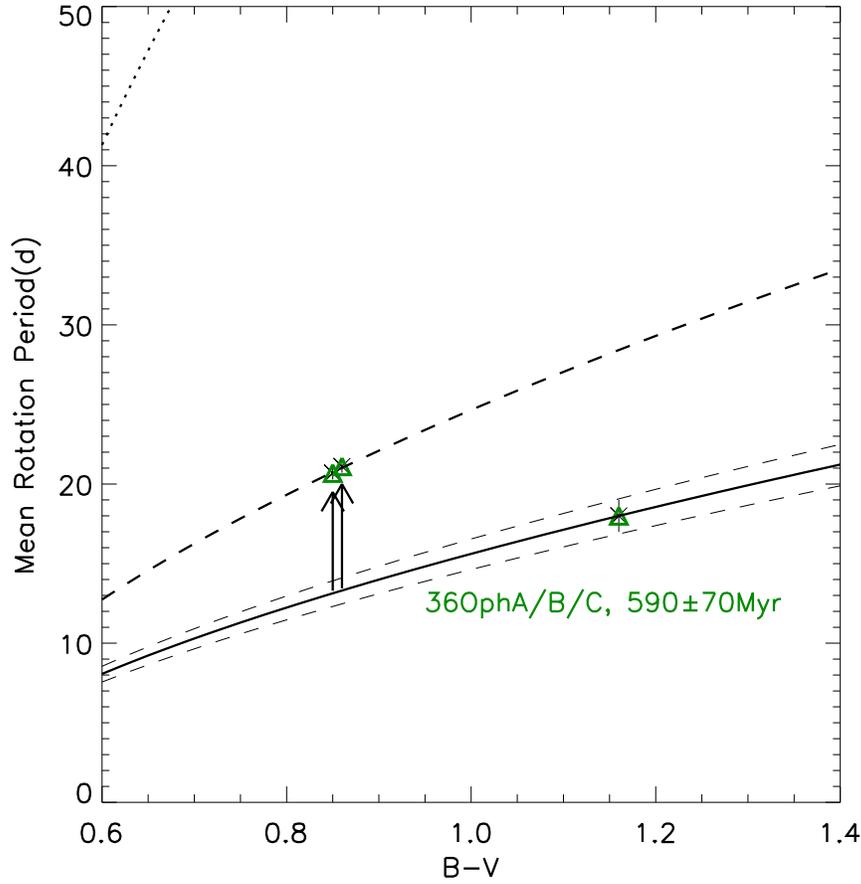}
\caption{Color-period diagram for the 36\,Oph\,ABC triple system.
Isochrones are drawn for ages of 590\,Myr (solid) and 1.43\,Gyr (thick dashed). 
The distant companion, C, gives the 590\,Myr age for the system. 
The error is indicated with thin dashed lines.
The A and B components appear to have interacted and spun down to $\sim$20d 
against a nominally expected period of $\sim$13d. 
The dotted line corresponds to the age of the universe.
\label{fig17}}
\end{figure}

\begin{figure} 
\includegraphics[scale=1.00]{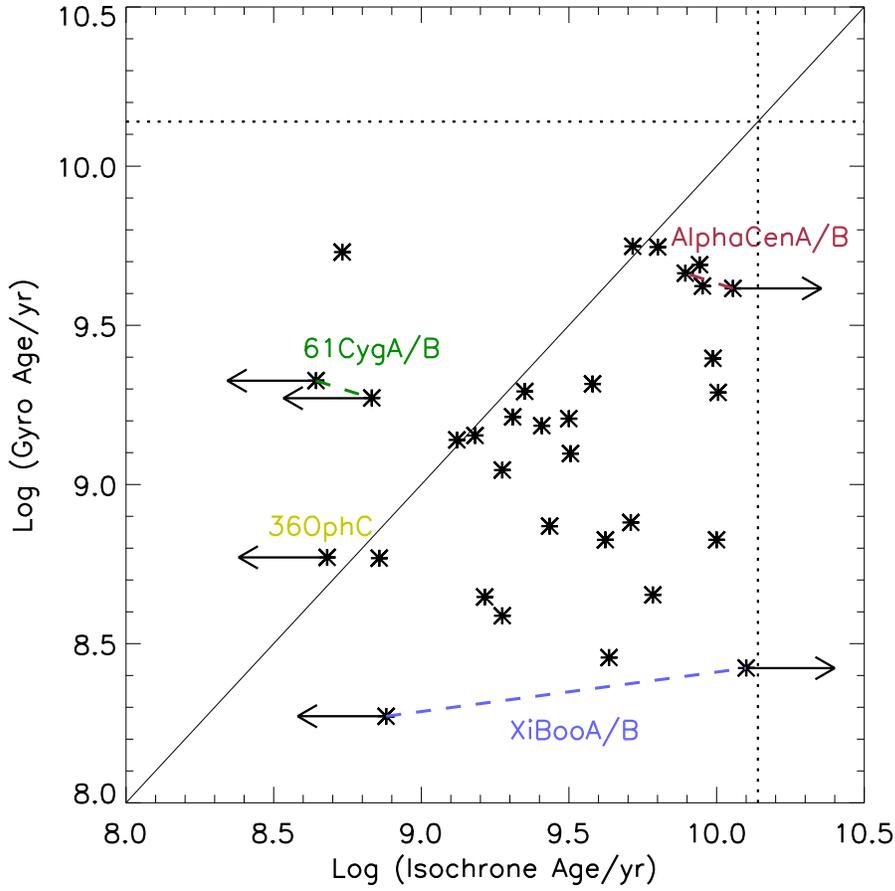}
\caption{Comparison of gyro and isochrone ages for the 26 Takeda et al. (2007)
stars with `well-defined' ages in common with the gyrochronology sample 
presented in this paper. The solid line denotes equality and the dotted lines 
the age of the universe. There is no strong correlation between the two ages, 
except that the median isochrone age is a factor of 2.7 times higher than 
the median gyro age. Takeda et al. (2007) stars with upper- or lower limits 
(arrows) are not plotted, except for the wide binaries (the components 
are connected by dashed lines) discussed in the text. It would appear that the
gyro ages supercede the isochrone ages for main sequence stars.
\label{fig18}}
\end{figure}








\end{document}